\documentclass[journal,twoside]{IEEEtran}
\usepackage{algorithm}
\usepackage{algpseudocode}
\usepackage{amsmath}
\usepackage{amssymb}
\usepackage{amsmath,epsfig,bm,subfigure}
\usepackage{multirow}
\usepackage{tabu}
\usepackage{makecell}
\usepackage{cite}
\usepackage{threeparttable}
\usepackage{url}
\usepackage{booktabs,bigstrut}

\ifCLASSINFOpdf
\else
\fi

\begin{document}
\title{A Convolutional Neural Network-Based Low Complexity Filter}
\author{Chao~Liu,~\IEEEmembership{Student Member,~IEEE}
        Heming~Sun,~\IEEEmembership{Member,~IEEE}
        Jiro~Katto,~\IEEEmembership{Member,~IEEE}
        Xiaoyang~Zeng,~\IEEEmembership{Member,~IEEE}
        and~Yibo~Fan

\thanks{This work was supported in part by the National Natural Science Foundation of China under Grant 61674041, in part by Alibaba Group through Alibaba Innovative Research (AIR) Program, in part by the STCSM under Grant 16XD1400300, in part by the pioneering project of academy for engineering and technology and Fudan-CIOMP joint fund, in part by the National Natural Science Foundation of China under Grant 61525401, in part by the Program of Shanghai Academic/Technology Research Leader under Grant 16XD1400300, in part by the Innovation Program of Shanghai Municipal Education Commission, in part by JST, PRESTO Grant Number JPMJPR19M5, Japan. (Corresponding author: Heming Sun and Yibo Fan.)}% <-this % stops a space

\thanks{C. Liu, Y. Fan and X. Zeng are with the State Key Laboratory of ASIC and System, Fudan University, Shanghai 200433, China (e-mail: chaoliu18@fudan.edu.cn; fanyibo@fudan.edu.cn; xyzeng@fudan.edu.cn).}% <-this % stops a space

\thanks{H. Sun is with the Waseda Research Institute for Science and Engineering,
Tokyo 169-8555, Japan and JST, PRESTO, 4-1-8 Honcho, Kawaguchi, Saitama,
332-0012, Japan (e-mail: hemingsun@aoni.waseda.jp).}

\thanks{J. Katto is with Waseda Research Institute for Science and Engineering, Tokyo 169-8555, Japan and the Graduate School of Fundamental Science and Engineering, Waseda University, Tokyo 169-8555, Japan (e-mail:katto@waseda.jp).}}

\maketitle
\begin{abstract}
%\boldmath
%For the task of removing artifacts in video compression, neural networks achieve better performance than traditional in-loop filtering methods. %1
%For the task of removing artifacts in video compression, many previous works have gotten more BD-rates saving with increasing the model complexity. %2

Convolutional Neural Network (CNN)-based filters have achieved significant performance in video artifacts reduction.
However, the high complexity of existing methods makes it difficult to be applied in real usage.
%Especially for the decoder, too much computational complexity, and excessive memory consumption make them difficult to be applied in practical applications.
In this paper, a CNN-based low complexity filter is proposed.
We utilize depth separable convolution (DSC) merged with the batch normalization (BN) as the backbone of our proposed CNN-based network.
Besides, a weight initialization method is proposed to enhance the training performance.
%Specifically, the knowledge from a pre-trained teacher model is used to improve the filtering performance without increasing complexity.
To solve the well known over smoothing problem for the inter frames, a frame-level residual mapping (RM) is presented. We analyze some of the mainstream methods like frame-level and block-level based filters quantitatively and build our CNN-based filter with frame-level control to avoid the extra complexity and artificial boundaries caused by block-level control. In addition, a novel module called RM is designed to restore the distortion from the learned residuals. As a result, we can effectively improve the generalization ability of the learning-based filter and reach an adaptive filtering effect.
Moreover, this module is flexible and can be combined with other learning-based filters.
The experimental results show that our proposed method achieves significant BD-rate reduction than H.265/HEVC.
It achieves about 1.2\% BD-rate reduction and 79.1\% decrease in FLOPs than VR-CNN. Finally, the measurement on H.266/VVC and ablation studies are also conducted to ensure the effectiveness of the proposed method.
% ablation study is conducted for the proposed method.
\end{abstract}


% \fi
\IEEEpeerreviewmaketitle

\section{Introduction}
\IEEEPARstart{T}{he} performance of video compression has been continuously improved with the development from H.264/AVC\cite{h264}, H.265/HEVC\cite{h265} to H.266/VVC\cite{h266}. These standards share a similar hybrid video coding framework, which adopts prediction \cite{lainema2012intra,lin2013motion}, transformation \cite{nguyen2013transform}, quantization \cite{crave2010robust}, and context adaptive binary arithmetic coding (CABAC)\cite{marpe2003context}.
%The frames in the video are split into small blocks with a flexible quadtree coding structure, and the prediction tools are utilized to reduce the spatial and temporal correlation between those blocks. By using the discrete transformation on the difference between the original and predicted blocks, the correlation within blocks is further removed. After transformation, those residues are quantized and sent to lossless entropy codec. The distortion is mainly produced by quantization, which drops the high-frequency components of the transformed domain to have a trade-off between the rate and distortion.
Owing to the modules like quantization and flexible partition, some unavoidable artifacts are produced and cause degradation of video quality, such as blocking effect, Gibbs effect, and ringing. To compensate for those artifacts, many advanced filtering tools are designed, for instance, de-blocking(DB\cite{norkin2012hevc}), sample adaptive offset(SAO\cite{fu2012sample}), and adaptive loop filter(ALF\cite{tsai2013adaptive}). These tools reduce the artifacts effectively with acceptable complexity.

In the past decades, the learning-based methods make great progress in both low-level and high-level computer vision tasks\cite{duan2019centernet,zhao2019object,liu2019recent,liu2019auto,hu2019meta,soh2019natural}, such as object detection\cite{duan2019centernet,zhao2019object}, semantic image segmentation\cite{liu2019recent,liu2019auto}, and super resolution\cite{hu2019meta,soh2019natural}.
By virtue of the powerful non-linear capability of learning-based tools, they also have been utilized to replace the existing modules in video coding and show great potential, for instance, intra prediction\cite{li2018fully,hu2019progressive,sun2020enhanced}, inter prediction\cite{liu2018one,zhao2019enhanced}, and entropy coding\cite{song2017neural,ma2019neural}.
%Those learning-based methods have achieved considerable BD-rate \cite{bdrate} reduction than original video coding modules.
Learning-based models, especially CNN, have achieved excellent performances for the in-loop filter of video coding \cite{IFCNN,VRCNN,dai2018cnn,liu2019dual,MMNN,RHCNN,SJNN,RSNN,jia2019content}.
%To adapt the variable sizes of transformation in H.265/HEVC,
Dai \textit{et al}. \cite{VRCNN,dai2018cnn} proposed VR-CNN, which adopts a variable filter size technique to have different receptive fields in one-layers and achieves excellent performance with relatively low complexity. Zhang \textit{et al.} \cite{RHCNN} proposed a 13-layer RHCNN for both intra and inter frames. The relatively deep network has a strong mapping capability to learn the difference between the original and the reconstructed inter frames. To further adapt to the image content, Jia \textit{et al.} \cite{jia2019content} designed a multi-model filtering mechanism and proposed a content-aware CNN with a discriminative network. This method uses the discriminative network to select the most suitable deep learning model for each region.

Most of the learning-based filters can achieve considerable BD-rate \cite{bdrate} savings than H.265/HEVC anchor. However, real-world applications often require lightweight models. High memory usage and computing resource consumption make it difficult to apply complex models to various hardware platforms. Therefore, designing a light network is essential to popularize learning-based in-loop filters. Considering this, some model compression methods that reduce the model complexity while maintaining performance are needed.
In recent years, some famous methods have been proposed, including lightweight layers\cite{sifre2014rigid,howard2017mobilenets}, knowledge transfer \cite{Hinton2015Distilling,zagoruyko2016paying,huang2017like}, low-bit quantization\cite{yang2019quantization,nagel2019data}, and network pruning\cite{molchanov2019importance,zhao2019variational}. DSC \cite{sifre2014rigid,howard2017mobilenets} is one of the famous lightweight layers. It preserves the essential features of standard convolution while greatly reducing the complexity by using grouping convolution\cite{sifre2014rigid}. In this paper, we build our learning-based filter with DSC instead of the standard convolution. And knowledge transfer is used to help the initialization of the trainable parameters without increasing the complexity.
%It trains the model by using the prior knowledge obtained from another pre-trained model.

%Yang \textit{et al.} \cite{SCNN} proposed a recurrent greedy approach for inter frames.
Besides the learning-based filter itself, we also need a lightweight mechanism for the filtering of inter frames. Some inter blocks fully inherit the texture from their reference blocks and have almost no residuals. If the learning-based filter is used for each frame, those blocks will be repeatedly filtered and cause over-smoothing in inter blocks\cite{jia2019content, RRCNN}. One solution to solve this problem is training a specific filter for inter frames \cite{RHCNN}. However, the coding of intra and inter frame share some of the same modules in H.265/HEVC like transformation, quantization, and block partitions. This means the learning-based filter trained with intra frames can also be used for inter frames to some extent.
Considering this, previous works \cite{dai2018cnn, jia2019content, STResNet,jvetCtu, RRCNN, SDLA} designed a syntax element control flag to indicate whether an inter CTU uses the learning-based filter or not. It chooses a selective filtering strategy for each CTU. For this strategy, we compare it with frame-level control in Section \ref{analysisCtuFrame} and found the CTU-level control may lead to artificial boundaries between the neighboring CTUs. So we propose to use the frame-level based filter to avoid unnecessary artificial boundaries. In order to improve the performance of frame-level based filtering, we propose a novel module called residual mapping (RM) in this paper.
%RM chooses the best filtering strength with the least square errors and achieves better video subjective and objective quality in experiments.

%Therefore, we design a novel low-complexity filtering algorithm for inter frames based on frame-level.
In summary, we propose a novel light CNN-based in-loop filter for both intra and inter frames based on \cite{sun2020image,liu2020learning}. Experimental results show this model achieves excellent performance in terms of both video quality and complexity. Specifically, our contributions are as follows.

\begin{itemize}
  \item A CNN-based lightweight in-loop filter is designed for H.265/HEVC. Low-complexity DSC merged with the BN is used as the backbone of this model.
      %We analyze the difference between DSC and standard convolution to construct our proposed model.
      Besides, we use attention transfer to pre-train it to help the initialization of parameters.
  \item For the filtering of inter frames, we analyze and build our CNN-filter based on frame-level to avoid the artificial boundaries caused by CTU-level. Besides, a novel post-processing module RM is proposed to improve the generalization ability of the frame-level based model and enhance the subjective and objective quality.
  \item We integrate the proposed method into HEVC and VVC reference software and significant performance has been achieved by our proposed method. Besides, we conduct some extensive experiments like ablation studies to prove the effectiveness of our proposed methods.
\end{itemize}

The following of this paper is organized as follows. In Section II, we present the related works, including the in-loop filter in video coding and the lightweight network design. Section III elaborates on the proposed network, including network structure and its loss function. Section IV focuses on the proposed RM module and provides an analysis of different control strategies. Experiment results and ablation studies are shown in Section V. In Section VI, we conclude this paper with future work.

\section{Related Works}
\subsection{In-loop Filters in Video Coding}
%To alleviate the artificial imprints caused by the lossy block-based transform compression framework, various in-loop filtering methods have been proposed in the past decades. In this subsection, we briefly review the previous work including DB, SAO, ALF, and learning-based filters.
\subsubsection{DB, SAO, and ALF}
DB, SAO, and ALF that are adopted in the latest video coding standard H.266/VVC\cite{h266} are aimed at removing the artifacts in video coding.
De-blocking \cite{norkin2012hevc} has been used to reduce the discontinuity at block boundaries since the publication of coding standard H.263+\cite{cote1998h}. Depend on the boundary strength and reconstructed average luminance level, DB chooses different coding parameters to filter the distorted boundaries.
Meanwhile, by classifying the reconstructed samples into various categories, SAO \cite{fu2012sample} gives each category a different offset to compensate for the error between the reconstructed and original pixels. Based on the Wiener filter, ALF \cite{tsai2013adaptive} tries different filter coefficients by minimizing the square error between the original and reconstructed pixels. The signal of the filter coefficient needs to be sent to the decoder side to ensure the consistency between encoder and decoder.
All these aforementioned filters can effectively alleviate the various artifacts in reconstructed images. However, there is still much room for improvement.

\subsubsection{Learning-based Filter}
Recently, the learning-based filters have far outperformed the DB, SAO, and ALF in terms of both objective and subjective quality. Different from SAO and ALF, they hardly need extra bits but can compensate for errors adaptively as well. Most of them are based on CNNs and have achieved great success in this field.
For the filtering of intra frames, Park \textit{et al.} \cite{IFCNN} first proposed a CNN-based in-loop filter IFCNN for video coding. Dai \textit{et al.} \cite{VRCNN} proposed VR-CNN as post-processing to replace DB and SAO in HEVC. Based on inception, Liu \textit{et al.} \cite{liu2019dual} proposed a CNN-based filter with 475,233 trainable parameters. Meanwhile, Kang \textit{et al.} \cite{MMNN} proposed a multi-modal/multi-scale neural network with up to 2,298,160 parameters. Considering the coding unit (CU) size information, He \textit{et al.} \cite{SJNN} proposed a partition-masked CNN with a dozen residual blocks.
Sun \textit{et al.} \cite{sun2020image} proposed a learning-based filter with ResNet\cite{resnet} for the VTM.
Liu \textit{et al.} \cite{liu2020learning} proposed a lightweight learning-based filter based on DSC.
Apart from what was mentioned above, Zhang \textit{et al.} \cite{RRCNN} proposed a residual convolution neural network with a recursive mechanism.

Different from the training the filter for intra samples, training the filter with inter samples need to consider the problem of repeated filtering \cite{jia2019content, SDLA}. Jia \textit{et al.} \cite{jia2019content} proposed a content-aware CNN based in-loop filtering method that applies multiple CNN models and a discriminative network in the H.265/HEVC. This discriminative network can be used to judge the degree of distortion of the current block and select the most appropriate filter for it.
However, the discriminative network requires additional complexity and memory usage, some researchers \cite{STResNet,dai2018cnn} proposed to use block-level syntax elements to replace it. This method requires extra bit consumption but gets a more accurate judgment on whether to use the learning-based filter. Similarly, some researchers \cite{IFCNN,wang2018dense} proposed to use frame-level syntax elements to control the filtering of inter frames.
Besides, complicated models \cite{STResNet,RHCNN,soh2018reduction} like spatial-temporal networks are also useful for solving this problem. Jia \textit{et al.} \cite{STResNet} proposed spatial-temporal residue network (STResNet) with CTU level control to suppress visual artifacts.
RHCNN that is trained for both intra and inter frames was proposed by Zhang \textit{et al.} \cite{RHCNN}.
Filtering in the decoder side \cite{RSNN,li2017cnn,zhang2020enhancing} can also solve the problem of repeated enhancement well. For example, DS-CNN was designed by Yao \textit{et al.} \cite{RSNN} to achieve quality enhancement as well. Li \textit{et al.} \cite{li2017cnn} adopted a 20-layers deep CNN to improve the filtering performance. Zhang \textit{et al.} \cite{zhang2020enhancing} proposed a post-processing network for VTM 4.0.1.

In summary, filtering in inter frames is more challenging than that of intra frames.
In most cases, the CNN-based in-loop filter with higher complexity can achieve better performance on intra frames. But for the filtering of inter frames, the existing methods have their own problems. For example, frame-level control may lead to an over-smoothing problem, CTU-level control will cause the additional artificial boundaries, the out-loop filters cannot use the filtered image as a reference, adding discriminative network and complex model may lead to over-complexity and impractical. Therefore, we should pay attention to a more effective method for this task.

\subsection{Lightweight Network Design}
%Although complex models have better performance, high memory usage and computing resource consumption make it difficult to apply them to various hardware platforms effectively. Therefore, lightweight neural network design has become one of the hottest points of research in both industry and academia.
\subsubsection{Depthwise Separable Convolution}
\begin{figure}[!tbp]
  \centering
  \includegraphics[scale=0.55]{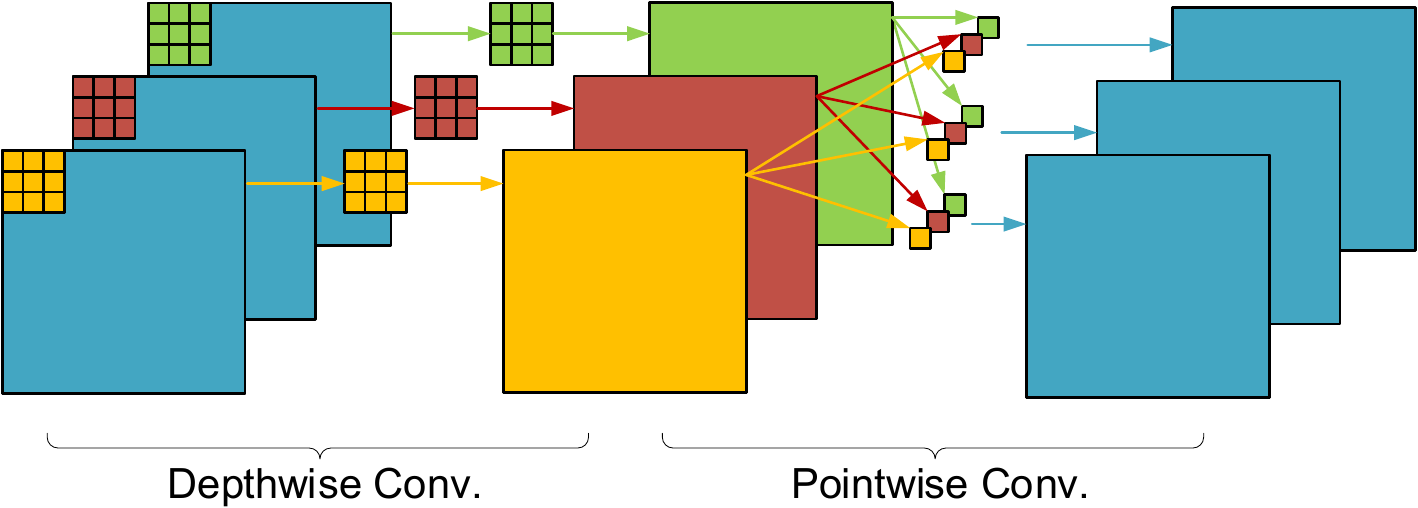}
  \caption{The depthwise separable convolution, where "Conv." indicates convolution.}\label{dsconv}
\end{figure}

As a novel neural network layer, DSC achieves great success in practical applications because of its low complexity. It is initially introduced in \cite{sifre2014rigid} and subsequently used in MobileNets \cite{howard2017mobilenets}.
As shown in Fig. \ref{dsconv}, DSC divides the calculation of standard convolution into two parts, depthwise convolution, and pointwise convolution. Different from standard convolution, depthwise convolution decompose the calculation of standard convolution into group convolution to reduce the complexity. Meanwhile, the pointwise convolution is the same as the standard convolution with kernel $1\times 1$. In other words, depthwise convolution is used to convolute the separate features whereas pointwise convolution is utilized to combine them to get the output feature maps. These two parts together form a complete DSC.

\subsubsection{Knowledge Distillation and Transfer}
Previous studies \cite{Hinton2015Distilling,huang2017like,zagoruyko2016paying} have shown that the "knowledge" in pre-trained models can be transferred to another model.
Hinton \textit{et al.} \cite{Hinton2015Distilling} propose a distillation method that uses a teacher model to get a "soft target", which helps a student model that has a similar structure perform better in the classification task. Besides softening the target in classification tasks, some other methods \cite{zagoruyko2016paying,huang2017like} use the intermediate representations of the pre-trained model to transfer the "knowledge". For example,
Zagoruyko \textit{et al.} \cite{zagoruyko2016paying} devise a method called attention transfer (AT) to get student model performance improved by letting it mimic the attention maps from a teacher model.
Meanwhile, Huang \textit{et al.} \cite{huang2017like} design a loss function by minimizing the maximum mean discrepancy (MMD) metric between the distributions of the teacher and the student model, where MMD is a distance metric for probability distributions \cite{gretton2012kernel}.
%The squared MMD of samples $\mathcal{X} =\{ x^i \}^N_{i=1}$ and $\mathcal{Y} =\{ y^j \}^M_{j=1}$ from distributions $\bm{p}$ and $\bm{q}$ can be formulated as:
%\begin{align}\label{mmd1}
%  &\mathcal{L}_{ MMD^2}(\mathcal{X},\mathcal{Y}) \notag \\
%  = &\| \frac{1}{N}\sum_{i=1}^{N}\phi(x^i) - \frac{1}{M}\sum_{j=1}^{M}\phi(y^j)\|^2_2 \\
%   = &\frac{1}{N^2}\sum_{i=1}^{N}\sum_{i'=1}^{N}\phi(x^i)^T\phi(x^{i'}) + \frac{1}{M^2}\sum_{j=1}^{M}\sum_{j'=1}^{M}\phi(y^j)^T\phi(y^{j'}) \notag  \\
%  & - \frac{2}{MN}\sum_{i=1}^{N}\sum_{j=1}^{M}\phi(x^i)^T\phi(y^j) 	\\
%  = &\frac{1}{N^2}\sum_{i=1}^{N}\sum_{i'=1}^{N}k(x^i,x^{i'}) + \frac{1}{M^2}\sum_{j=1}^{M}\sum_{j'=1}^{M}k(y^j,y^{j'}) \notag  \\
%  & - \frac{2}{MN}\sum_{i=1}^{N}\sum_{j=1}^{M}k(x^i,y^j)
%\end{align}
%where $\phi(\cdot)$ is a explicit mapping function, $N$ and $M$ indicate the number of samples from distribution $\bm{p}$ and $\bm{q}$, and $k(x,y)$ is a kernel function, which projects the sample into a higher dimensional feature space. With specific kernel, we can quantitatively calculate the distance between $\bm{p}$ and $\bm{q}$. Minimizing the loss function $\mathcal{L}_{ MMD^2}(\mathcal{X},\mathcal{Y})$ in learning-based methods is equivalent to minimize this distance.

\section{Proposed CNN-based Filter}
\begin{figure*}[!tbp]
  \centering
  \includegraphics[scale=0.65]{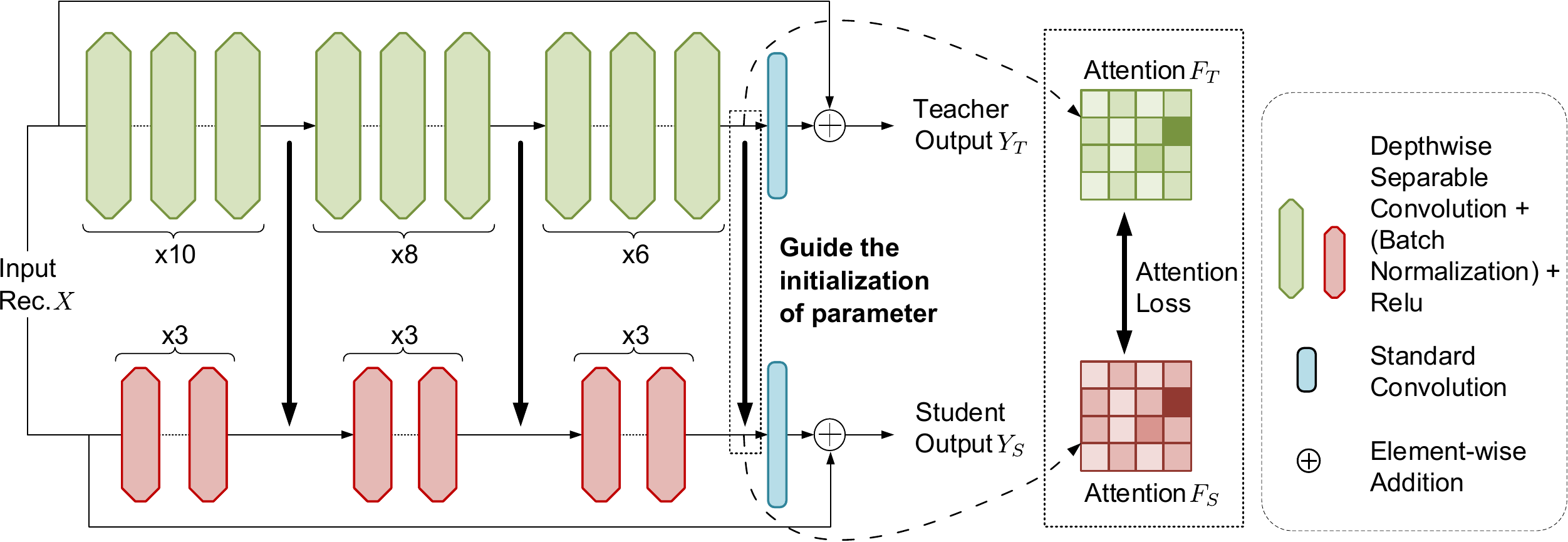}
  \caption{The architecture of teacher model and the proposed model, where "Rec." indicates "reconstructed pixels". The top-right and bottom-right are the teacher model and the proposed student model, respectively. The rectangle on the right implies the knowledge transfer.}\label{model}
\end{figure*}

\subsection{Network Structure}
As shown in Fig. \ref{model}, we design a network structure that functions on both the teacher and the proposed model.
This structure is composed of convolution, BN layer, and activation ReLU \cite{ReLU}. The backbone of this structure is $K$ layers of DSC with dozens of feature maps $F$.
The input to this structure is the HM reconstruction without filtering and the output is the filtered reconstructed samples. The last part is a standard convolution with only 1 feature map. And we add the reconstruction samples to the output inspired by residual learning \cite{resnet}.
The depthwise and the standard convolution kernel are both $3\times 3$. Every convolution is followed by the ReLU except for the last one.
The reason why choose ReLU instead of other advanced activation functions is that ReLU has a lower complexity while a considerable nonlinearity.
%It is also worth noting that the last convolution of the proposed model is standard convolution rather than DSC. This is because standard convolution uses less calculation than DSC when there is only one convolution output channel.
In our implementation, the values of $K$ and $F$ are 24 and 64 for the teacher model, 9 and 32 for the proposed model. The description of the parameters in the proposed model is shown in Table \ref{paranum}.
%\begin{figure}[!tbp]
%  \centering
%  \includegraphics[scale=0.55]{./dsconv.pdf}
%  \caption{The depthwise separable convolution.}\label{dsconv}
%\end{figure}

We use the BN layer in the training phase, this layer could improve the back-propagation of the gradients. What's more, both BN and convolution are linear computations for the tensors in the proposed model. Therefore, the BN can be merged into the convolution to further reduce the computational during the inference phase. As shown in (\ref{dwconv}), depthwise convolution output $\bm{\chi}_{dwConv}$ can be formulated as:
\begin{equation}\label{dwconv}
  \bm{\chi}_{dwConv}  = \bm{w}_{dwConv}*  \bm{\chi}
\end{equation}
where $*$ indicates the convolution operation, $\bm{w}_{dwConv}$ is the kernel and $\bm{\chi}$ is the depthwise convolution input.
Similarly, the piecewise convolution output $\bm{\chi}_{pwConv}$ can be written as:
\begin{equation}\label{pwconv}
    \bm{\chi}_{pwConv}  = \bm{w}_{pwConv}* \bm{\chi}_{dwConv} + \bm{b}_{pwConv}
\end{equation}
where $\bm{w}_{pwConv}$ and $\bm{b}_{pwConv}$ denote the kernel and bias.
It is noticeable in (\ref{dwconv}) that the convolution of depthwise convolution has no bias, this is because the bias $\bm{b}_{dwConv}$ can be merged into $\bm{b}_{pwConv}$ when there is no activation between depthwise and pointwise convolution.
After convolution, the output of BN can be obtained by (\ref{bnb}). (The reason why we use $*$ operation here is because actually the calculation of BN is equivalent to that of the depthwise convolution by simplification)
\begin{equation}\label{bnb}
  \bm{\chi}_{bn} = \bm{\gamma} * \left( \frac{\bm{\chi}_{pwConv}-\bm{mean}}{\sqrt{\bm{var}+\bm{\epsilon}}} \right)+\bm{\beta}
\end{equation}
Substituting (\ref{pwconv}) into (\ref{bnb}), we obtain (\ref{bnn}) as follows:
\begin{equation}\label{bnn}
  \bm{\chi}_{bn} = \bm{\widehat{w}}_{pwConv}* \bm{\chi}_{dwConv} + \bm{\widehat{b}}_{pwConv}
\end{equation}
where $\bm{\widehat{w}}_{pwConv}$ and $\bm{\widehat{b}}_{pwConv}$ in (\ref{bnn}) are:
\begin{align}
  \bm{\widehat{w}}_{pwConv} &= \frac{\bm{\gamma} * \bm{w}_{pwConv}}{\sqrt{\bm{var}+\bm{\epsilon}}} \label{wpwConv} \\
  \bm{\widehat{b}}_{pwConv} &= \frac{\bm{\gamma} * (\bm{b}_{pwConv}-\bm{mean})}{\sqrt{\bm{var}+\bm{\epsilon}}}+\bm{\beta} \label{bpwConv}
\end{align}
In (\ref{wpwConv}) and (\ref{bpwConv}), $\bm{\gamma}$ and $\bm{\beta}$ are trainable parameters of BN, $\bm{mean}$ and $\bm{var}$ are non-trainable parameters of BN. Hyper-parameter $\bm{\epsilon}$ represents a positive number that prevents division zero errors. In the inference phase, we use the $\bm{\widehat{w}}_{pwConv}$ and $\bm{\widehat{b}}_{pwConv}$ to replace the weight $\bm{w}_{pwConv}$ and bias $\bm{b}_{pwConv}$ in depthwise convolution, thus merging the BN into the DSC and reducing the model complexity.

\begin{table}[!tbp]
\begin{threeparttable}
  \centering
  \caption{Description of the Parameters in the Proposed Model}\label{paranum}%添加标题 设置标签
    \setlength{\tabcolsep}{1mm}{
  \begin{tabular}{l|c|c|c|c|c}
    \Xhline{1.0pt}
   Index  &Block1&Block2&Block3&Std Conv.\tnote{a} &Sum \bigstrut\\
   \hline
   Parameters &$73+2\times1,344$ &$3\times1,344$ &$3\times1,344$ &$289$ &11,114 \bigstrut\\
    \Xhline{1.0pt}
    %\bottomrule
  \end{tabular}}
      \begin{tablenotes}
        \item[a] Standard Convolution.
    \end{tablenotes}
\end{threeparttable}
\end{table}

\subsection{Standard Convolution of the Proposed Structure}
In this subsection, the last part of the proposed structure is detailed. Because the standard convolution uses fewer calculations than DSC when the number of convolution output channels is only one. It is worth noting that the last convolution of the proposed model is standard convolution, which isn't consistent with the backbone of the proposed model.
The DSC consists of two steps, including depthwise convolution and pointwise convolution. The depthwise convolution is the simplification of the standard convolution to reduce the amount of computation while preserving the ability to convolve the input feature maps. Meanwhile, the pointwise convolution is equivalent to the standard convolution with $1\times 1$ kernel, it is utilized to fuse the different depthwise convolution output. According to their computing methods, the ratio $r$ of the calculation of the DSC to that of the standard convolution is calculated as:
\begin{equation}\label{ratio}
  r= \frac{K_W K_H  C_I  W H +C_I C_O  W H }{K_W K_H  C_I C_O  W H}=\frac{1}{C_O}+\frac{1}{K_WK_H}
\end{equation}
where $W $, $ H$ is the width and height of the input frame, respectively. $K_W $, $K_H $ is the width and height of the convolution kernel, respectively. $C_I $, $C_O $ are the number of feature maps for the convolution input and output, respectively. In our proposed model, $C_O=1$ and $K_W=3, K_H=3$. So $r =\frac{1}{C_O}+\frac{1}{K_WK_H}=\frac{10}{9}$, which is bigger than $1$. This represents DSC consumes more computing sources than standard convolution.
The extra calculation is caused by pointwise convolution, which is utilized to combine feature maps. However, the standard convolution also can combine features, which indicates the extra calculation of pointwise convolution is meaningless. Therefore, we choose the standard convolution at the end of the model to avoid meaningless calculations.

\subsection{Proposed Initialization and Training Scheme}
In this subsection, we will introduce the training process and loss functions of the proposed network.
In most cases, a suitable initialization of parameters can help the model better converge to the minimum. Inspired by transfer learning, a pre-trained teacher model is used to guide the initialization of the parameters in the proposed model. By using such initialization, we hope the proposed model can obtain the output similar to that of the teacher model before the real training begins.
The pre-trained model uses the mean square errors (MSE) loss between the output $Y_T$ of teacher model and the original pixels $Y_O$.
\begin{equation}\label{lt}
  \mathcal{L}_T = \frac{1}{N}\sum_{i=1}^{N}\|Y_T^i-Y_O^i\|^2_2
\end{equation}
After the training of the teacher model, we use the intermediate outputs of it to guide the proposed model on parameter initialization. This process is denoted by the bold lines in Fig. \ref{model}. Because the vanishing of gradients may lead to insufficient training of shallow layers, the teacher model is divided into differently-sized blocks to produce the intermediate hints. The metric of the distance between teacher and the proposed student model tries two forms, including MMD \cite{huang2017like} and attention loss \cite{zagoruyko2016paying}. The loss function $\mathcal{L}_{MMD^2}(F_T,F_S)$ with linear kernel function ($k(x, y)=x^Ty$) could be written as follows:
\begin{equation}\label{mmdloss}
  \mathcal{L}_{MMD^2}(F_T,F_S) = \|\frac{1}{C_T}\sum_{i=1}^{C_T}\frac{f_T^i}{\|f_T^i\|_2}
  -\frac{1}{C_S}\sum_{j=1}^{C_S}\frac{f_S^j}{\|f_S^j\|_2}\|^2_2
\end{equation}
where $F$ represents the attention map, $f$ indicates a single feature map, $C$ is the number of feature maps, and the subscript $T$ and $S$ identify the teacher and student model.
Meanwhile, the loss function $\mathcal{L}_{AT}(F_T,F_S)$ of attention transfer (AT)\cite{zagoruyko2016paying} could be written as follows:
\begin{equation}\label{at}
  \mathcal{L}_{AT}(F_T,F_S) = \|\frac{\sum_{i=1}^{C_T}|f_T^i|^p}{\|\sum_{i=1}^{C_T}|f_T^i|^p\|_2}
  -\frac{\sum_{j=1}^{C_S}|f_S^j|^p}{\|\sum_{j=1}^{C_S}|f_S^j|^p\|_2}\|^2_2
\end{equation}
We set p to 2 in our implementation, because these two methods are similar except for their normalization methods when $p=1$\cite{huang2017like}.
%AT is to calculate the hint and then normalize the hint, and MMD is to normalize the feature maps first, then average them to get the hint.
After the initialization, we start the real training process of using MSE $\mathcal{L}_S$ in (\ref{ls}) to train the proposed model, where $Y_S$ indicates the output of the proposed model.
\begin{equation}\label{ls}
  \mathcal{L}_S =\frac{1}{N}\sum_{i=1}^{N} \|Y_S^i-Y_O^i\|^2_2
\end{equation}

In summary, the whole process can be divided into the following steps.
\begin{algorithm}[htb]
\normalsize
  \caption{ The process of building the trained proposed model.}
  \label{alg:Framwork}
  \begin{algorithmic}[1]
    \Require
      The dataset pair of HM reconstruction samples $X$ and original samples $Y_O$;
    \Ensure
      The trained proposed model;
    \State Constructing the teacher model $T$ and training it for $n_1$ epochs with MSE $\mathcal{L}_T $;
    \label{code:fram:ConstructingT}
    \State Extracting the attention maps $F_T$ from the trained $T$;
    \label{code:fram:Extracting}
    \State Constructing the student model $S$ with BN and training it for $n_2$ epochs with $\mathcal{L}_{AT}(F_T,F_S)$ or $\mathcal{L}_{MMD^2}(F_T,F_S)$;
    \label{code:fram:ConstructingSB}
    \State Training $S$ with MSE $\mathcal{L}_S$ for $n_3$ epochs;
    \label{code:fram:Training}
    \State Calculating the $\bm{\widehat{w}}_{pwConv}$ and $\bm{\widehat{b}}_{pwConv}$ for $S$;
    \label{code:fram:Calculating}
    \State Removing the BN from $S$;
    \label{code:fram:classify}
    \State Using the $\bm{\widehat{w}}_{pwConv}$ and $\bm{\widehat{b}}_{pwConv}$ to replace the weight $\bm{w}_{pwConv}$ and bias $\bm{b}_{pwConv}$ in depthwise convolution of $S$;
    \label{code:fram:select} \\
    \Return $S$;
  \end{algorithmic}
\end{algorithm}

%\begin{enumerate}
%    \item Using MSE \ref{lt} between the teacher network output and original pixels to train the teacher model.
%    \item To initialize the student model by using the hint produced by the well-trained teacher model to train it.
%    \item Using MSE \ref{ls} between the student network output and original pixels to train the student model.
%    \item Fusing the BN layers into the pointwise convolution in the student model to get the final model.
%\end{enumerate}

\section{Proposed Residual Mapping for the CNN-based Filtering}

\subsection{Analysis of CTU-level and Frame-level Control}\label{analysisCtuFrame}

\begin{figure}[tbp]
\centering
\subfigure[Convolution with valid padding]{
\begin{minipage}[t]{\linewidth}
\centering
\includegraphics[scale=0.9]{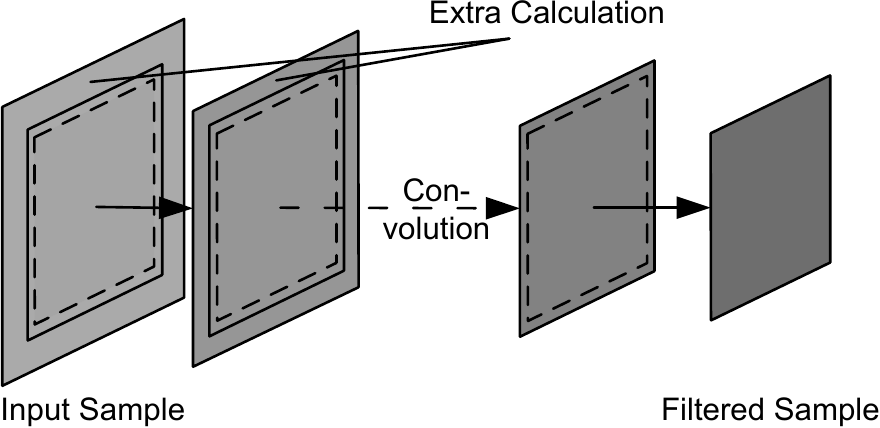}
\end{minipage}%
}%

\subfigure[Convolution with same padding]{
\begin{minipage}[t]{\linewidth}
\centering
\includegraphics[scale=0.9]{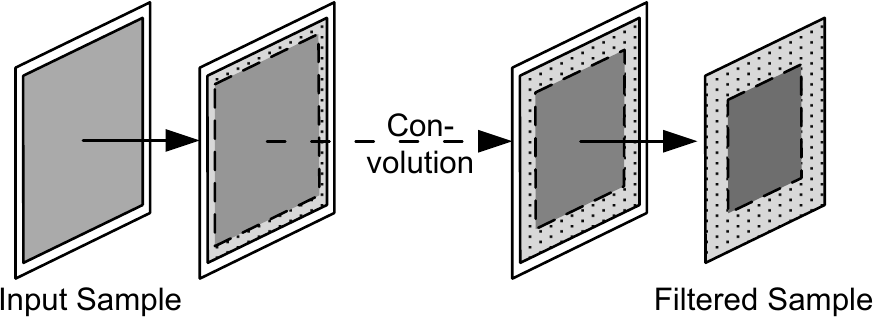}
\end{minipage}%
}

\subfigure{
\begin{minipage}[t]{\linewidth}
\centering
\includegraphics[scale=0.8]{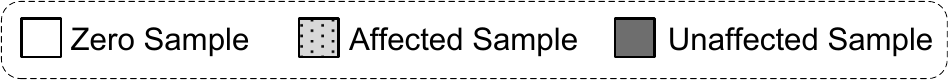}
\end{minipage}%
}%

\centering
\caption{The diagrams of convolution with different pad methods.}\label{padway}
\end{figure}

From the size of filtered samples, filtering methods can be divided into CTU-level (block-level) and frame-level.
Compared with CTU-level control, there are two main advantages of frame-level control in CNN-based filter design, including the required computational resource and the video quality. In this subsection, the difference is analyzed from the perspectives of the padding methods and the filter kernels.
\begin{table}[!tbp]
  \centering
  \caption{Complexity Comparison of CTU-level Control Between Valid Padding and Same Padding }\label{cp_pad}%添加标题 设置标签
  \begin{threeparttable}[b]
   \setlength{\tabcolsep}{1mm}{
  \begin{tabular}{l|c|c|c|c|c|c}
    \Xhline{1.0pt}
   Items &\multicolumn{2}{c|}{RHCNN\cite{RHCNN}} &\multicolumn{2}{c|}{Jia \textit{et al.} \cite{jia2019content}} & \multicolumn{2}{c}{VR-CNN\cite{VRCNN}}   \bigstrut \\
   % \cline{2-7}
   \hline
    Padding type  &Valid&Same&Valid&Same&Valid&Same \bigstrut\\
  % \hline
  % ClassA &  -5.50\%&130.9\%&2539.3\%&-7.50\%&105.4\%&737.3\%\\
   \hline \hline
    Flops\tnote{a} (G)&16.21&10.89&2.02&1.49&0.25&0.22 \bigstrut\\
    \hline
    Madd\tnote{b} (G)&32.38&21.76&4.04&2.97&0.49&0.44  \bigstrut\\
   \hline
   Memory\tnote{c} (MB)  &91.06&60.11&25.43&18.02&5.84&5.02\bigstrut\\
   \hline
   MemR+W\tnote{d} (MB)&193.05&130.36&55.09&39.42&13.88&11.99 \bigstrut\\
    %\hline
    %Parameters &\multicolumn{2}{c|}{362,753}&\multicolumn{2}{c|}{54,512}&\multicolumn{2}{c}{\textbf{11,114}} \\
    \Xhline{1.0pt}
    %\bottomrule
  \end{tabular}}
    \begin{tablenotes}
        \item[a] Theoretical amount of floating point arithmetics.
        \item[b] Theoretical amount of multiply-adds.
        \item[c] Memory useage.
        \item[d] Memory read /write.
    \end{tablenotes}
  \end{threeparttable}
\end{table}

\begin{figure}[tbp]
\centering
\subfigure[CTU-level control]{
\begin{minipage}[t]{.475\linewidth}
\centering
\includegraphics[scale=0.58]{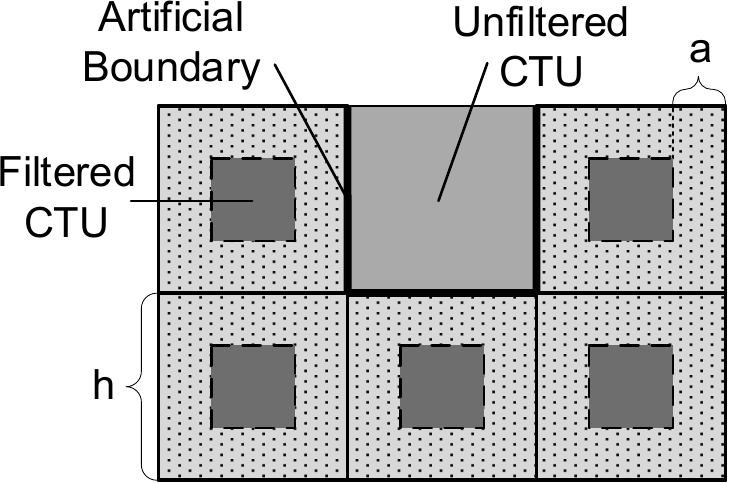}
\end{minipage}%
}
\subfigure[Frame-level control]{
\begin{minipage}[t]{.475\linewidth}
\centering
\includegraphics[scale=0.58]{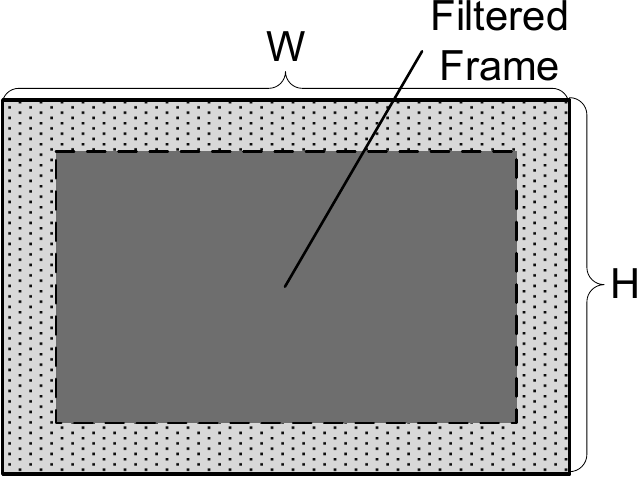}
\end{minipage}%
}%
\centering
\caption{The diagrams of convolution with different control methods.}\label{padcontrol}
\end{figure}

Firstly, to keep the input frames size unchanged, the CNN-based filter needs to pad the boundaries of input with some samples.
There are usually two padding ways, including valid padding (padded with reconstructed samples) and same padding (padded with zero samples).
In one case, if the CTUs are padded with reconstructed pixels to maintain the same accuracy as frame-level filtering, most of the networks need to pad the input block with plenty of pixels and require considerable calculation. Fig. \ref{padway} intuitively shows the difference in the amount of calculation between valid and same padding.
The quantitative calculations\cite{OpCounter} are illustrated in Table \ref{cp_pad} (we assume that both of their output sizes of filtered samples are $64\times 64$), it can be found that the valid padding (see "Valid" columns) of works \cite{VRCNN, RHCNN,jia2019content} all have considerable complexity increasing than same padding (see "Same" columns).
%In other words, choosing valid padding in lightweight convolutional network-based in-loop filter design consumes more computational resources.
In the other case, if the same padding is selected, it will cause calculation errors around the boundaries as shown in Fig. \ref{padcontrol}.
We assume that the size of the block control is $ h \times h $, and the width of the boundary area affected by the pad is $a$. The proportion $p_{fc}$ of affected pixels under frame-control is calculated as follows:
\begin{equation}\label{perimeter1}
  p_{fc}=1 - \frac{(W-2a)(H-2a)}{WH}=\frac{2a(W+H-2a)}{WH}
\end{equation}
Similarly, the proportion $p_{bc}$ of affected pixels under block control can be approximated as follows. (No incomplete CTU are considered)
\begin{equation}\label{perimeter2}
  p_{bc}=\frac{4a(h-a)}{h^2}\approx \frac{4a}{h}
\end{equation}
It can be found from (\ref{perimeter2}) that the area affected by same padding is approximately proportional to the perimeter of the filtered samples.
Therefore, the frame-level control with a higher area-to-perimeter ratio is less affected than block-level control.
According to (\ref{perimeter1}) and (\ref{perimeter2}), it can be obtained that for the HEVC test sequence, the same-padding of our network will affect an average of 45\% of the pixels under CTU-level control, whereas that of frame-level control is only 3\%.
Therefore, choosing frame-level control lays a solid foundation for the application of the CNN-based filter.

Secondly, the frames filtered by frame-level control has the property of integrity. Frame-level control uses the same kernel for filtering of the entire frame whereas CTU-level control may use the different kernels for two consecutive CTUs, which may lead to some artificial errors in the boundaries. As shown in Fig. \ref{padcontrol}, two consecutive CTUs with different filtering strategies have some errors along the boundaries because of the different kernels used in the filtering. Especially for the condition that one of the CTUs uses the learning-based filter while the other one doesn't.
This further demonstrates the advantages of frame-level control.

%\begin{table}[tbp]
%  \centering
%  \caption{The Required Bits and Proportion of Affected Pixels for Each Frame}\label{padtable}%添加标题 设置标签
%  \setlength{\tabcolsep}{1mm}{
%  \begin{tabular}{c|c|c|c}
%    \Xhline{1.0pt}
%   Sequences  & Resolution&
%   CTU-level&Frame-level \bigstrut\\
%   \hline \hline
%   ClassA & $2560\times1600$& 47.27\%& 1.01\%\bigstrut\\
%   \hline
%   ClassB & $1920\times1080$& 46.04\%& 1.44\%\bigstrut\\
%   \hline
%   ClassC & $832\times480$& 45.04\%& 3.26\%\bigstrut\\
%   \hline
%   ClassD & $416\times240$& 40.53\% &6.47\% \bigstrut\\
%   \hline
%   ClassE & $1280\times720$& 46.22\%& 2.16\%\bigstrut\\
%   \hline
%   \multicolumn{2}{c|}{Aveage} &  45.02\% &2.87\%\bigstrut\\
%
%    \Xhline{1.0pt}
%    %\bottomrule
%  \end{tabular}}
%\end{table}

In summary, for the design of lightweight CNN-based filters, the frame-level control has some advantages over block-level control.
On the one hand, compared with frame-level control, CTU-level control leads to calculation cost with the same padding or calculation error with the valid padding. On the other hand, frame-level control has the property of integrity and it brings better subjective quality.
To reduce the padding error brought by the multi-layer neural network and complexity, we built our CNN-based in-loop filters on a frame-level control. However, the ability to directly use frame-based control is weak because it only has two states of using or not using the filter, we need some added methods to improve its performance.

%This seems to be different from traditional filters such as SAO, ALF, and other block-based filters. The main reason for this difference is that the design of

\begin{figure}[tbp]
\centering
\subfigure[Org. frame]{
\begin{minipage}[t]{0.32\linewidth}
\centering
\includegraphics[width=2.7cm]{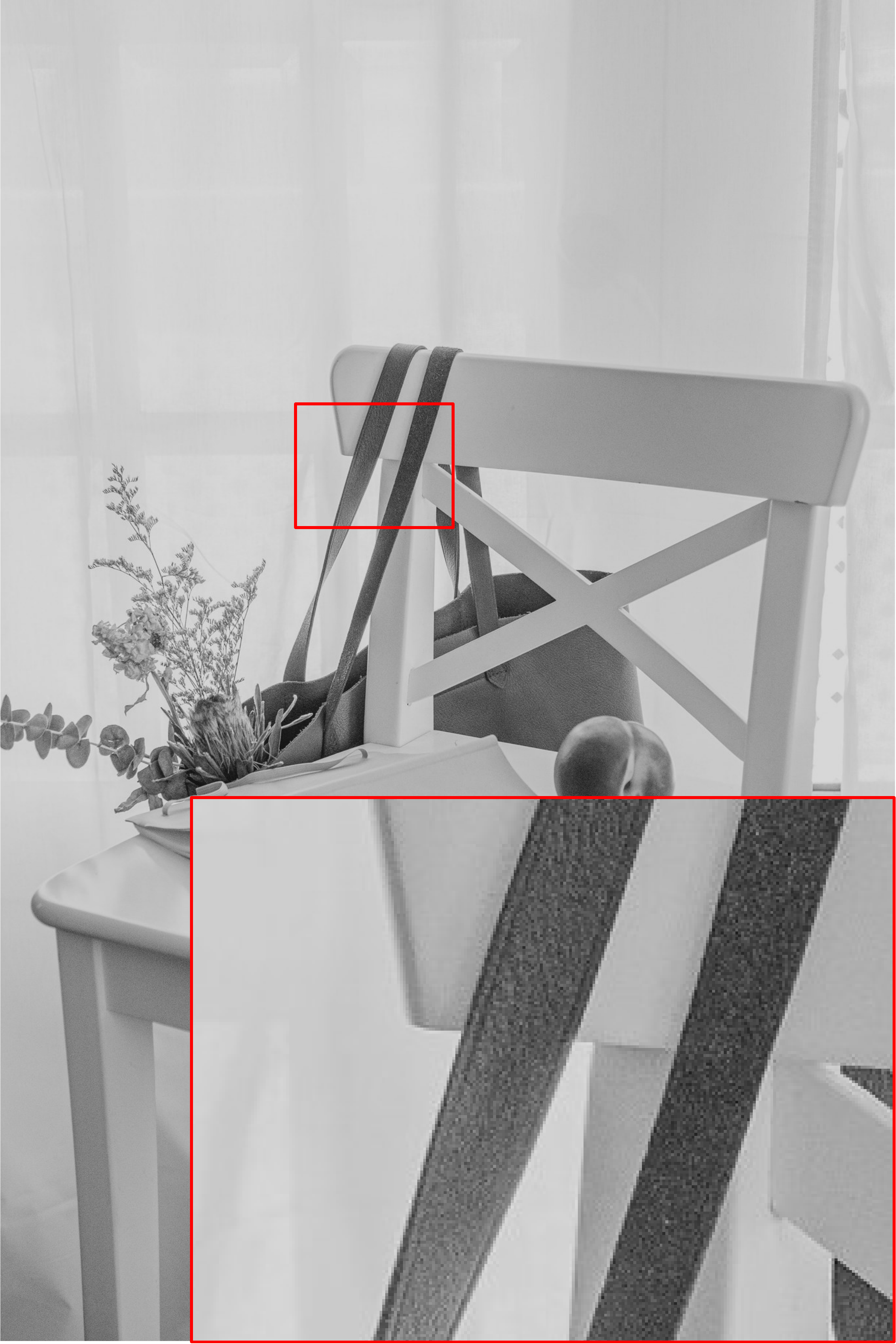}
\end{minipage}%
}%
\subfigure[Distortion]{
\begin{minipage}[t]{0.32\linewidth}
\centering
\includegraphics[width=2.7cm]{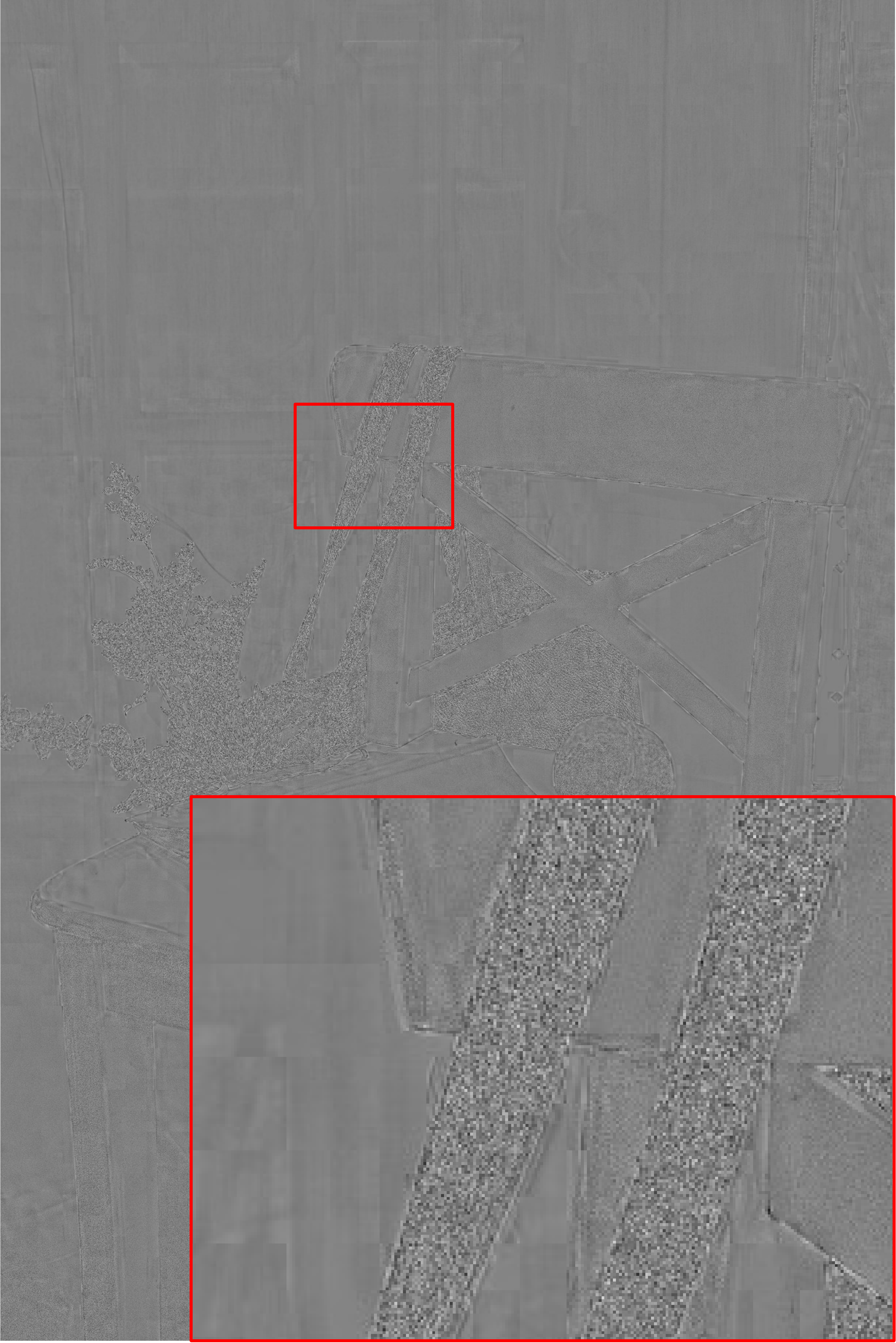}
\end{minipage}%
}%
\subfigure[Learned residual]{
\begin{minipage}[t]{0.32\linewidth}
\centering
\includegraphics[width=2.7cm]{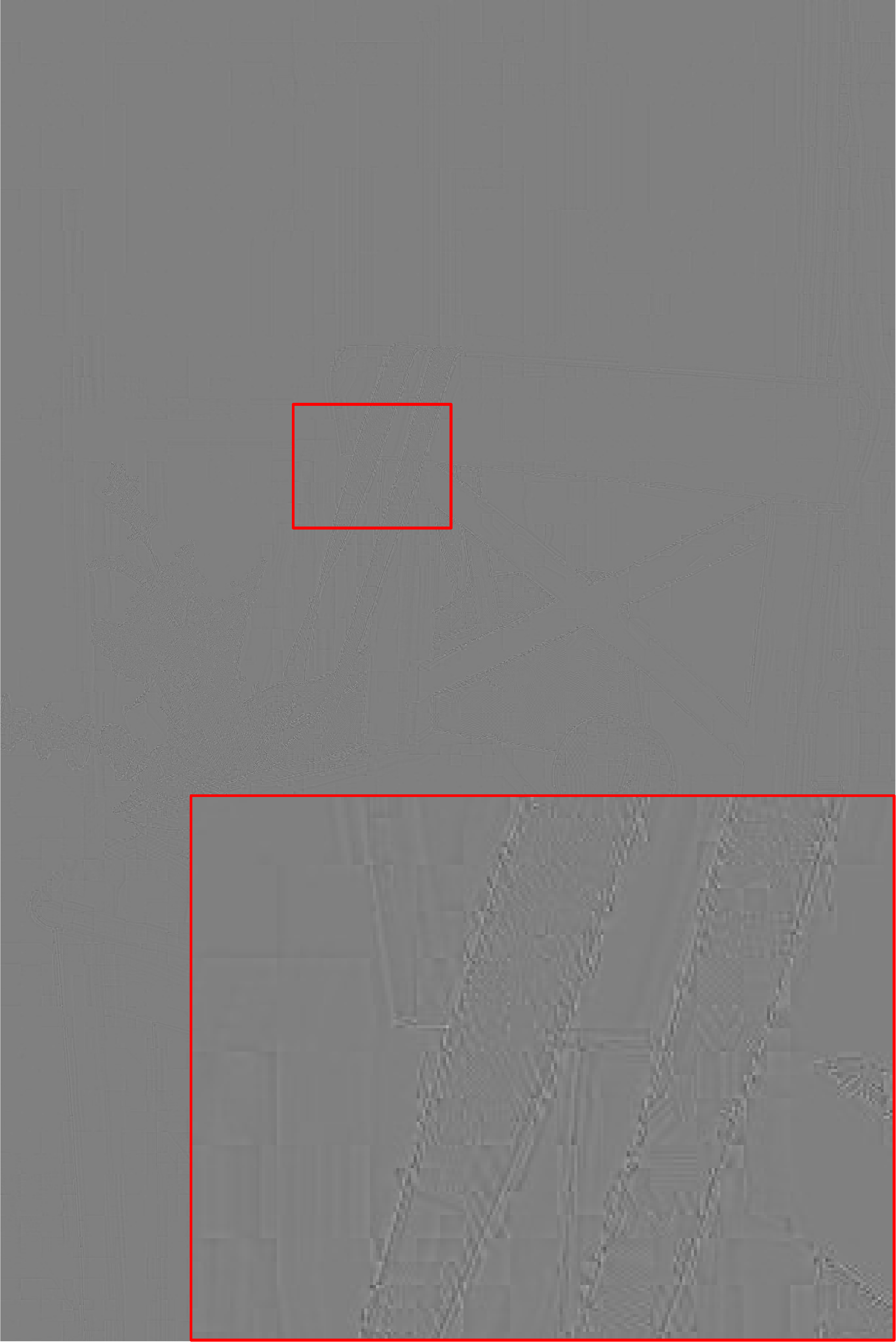}
\end{minipage}%
}%
\centering
\caption{A frame from CLIC dataset\cite{clic} is coded with HM-16.16 and QP 37. The original frame, the distortion and the learned residual of this frame are shown in (a), (b) and (c).}\label{odl}
\end{figure}

\subsection{Residual Mapping}
%When using the learning-based in-loop filter in H.265/HEVC, the relative increase in decoder complexity is significantly greater than that of the encoder.
%This is because the encoder needs to traverse different combinations of partitions and mode predictions while the decoder doesn't.
%大部分的训练好的神经网络都是对某个训练集的拟合，由于训练数据的分布往往很复杂，所以实际上训练是是对数据集的trade-off。对于某一张特定的图片而言，训练好的神经网络滤波器可能或多或少会欠滤波或者过滤波导致一定的失真或者过模糊。如果将使用intra样本训练好的神经网络用于inter样本进行滤波，这种现象则会更加严重因为其数据集分布存在一定的区别。

\begin{figure}[!tbp]
  \centering
  \includegraphics[width=8.3cm]{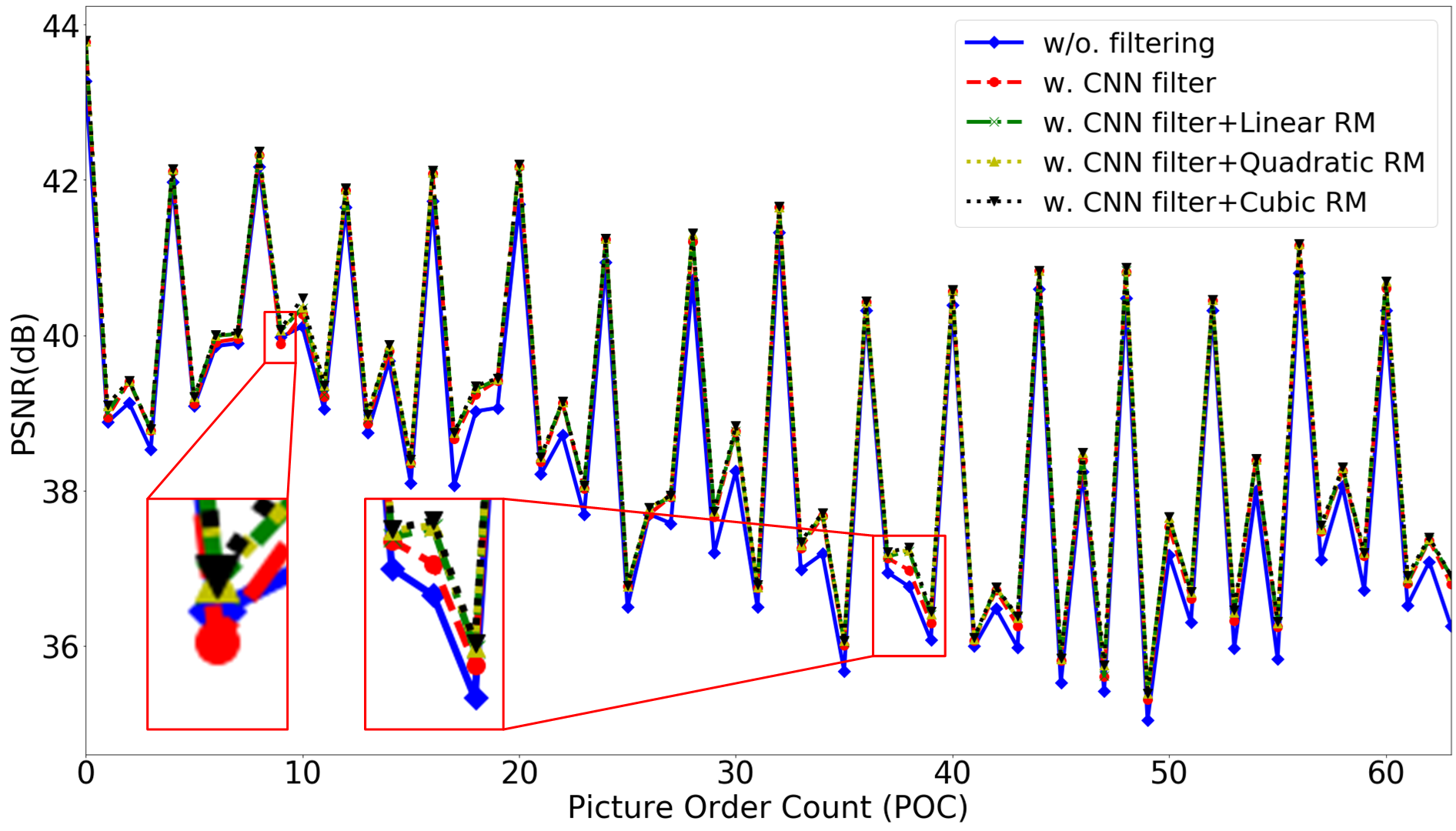}
  \caption{The comparison of different filtering mechanisms (``RaceHorses\_416x240", qp22, LDP configuration). Linear, quadratic, and cubic represent the mapping function of linear, quadratic, and cubic functions, respectively. We can find in the red box that the performance of using CNN-filter directly is not satisfactory, and even leads to a decrease in PSNR.}\label{rm_fig}
\end{figure}

In this subsection, a novel post-processing module RM is proposed to improve the performance of the frame-level control based CNN filter. It can effectively improve the over-smoothing problem \cite{jia2019content, SDLA} of inter frame. Besides, we found that it also has a considerable improvement to intra frames in Section \ref{rm_intra_ab}.
Most of the trained neural networks are fitting to a certain training set. Since the distribution of training data is often very complicated, the training is actually a trade-off of the data set. For a specific image, the trained filter may be under-fitted or over-fitted. This may cause distortion or blur for a learning-based filter. What's more, if we want to use the neural network trained with intra samples for the filtering of inter samples, this phenomenon will be more serious because of the difference in the distribution of the intra and inter datasets. With this in mind, we proposed to use a parametric RM after the learning-based filter, which is some sort of non-parametric filter, to improve its generalization ability.
Inspired by the potential correlation of distortion and learned residual shown in Fig. \ref{odl}, we handle this filter from the perspective that of restoring distortion from the learning-based filtered residual, which is equivalent to improving the quality of the distorted frames. The distortion $R_O$ is defined as the difference between the original samples $Y_O$ and reconstruction of de-blocking $X$ :
\begin{equation}\label{RO}
  R_O = Y_O - X
\end{equation}
Similarly, the learned residual $R_S$ is defines as the difference between the output of learning-based filter and $X$.
\begin{equation}\label{RS}
  R_S = Y_S - X
\end{equation}
A function $f_{\lambda}(\cdot)$ with parameters $\lambda$ is designed as the parametric filter to map $R_S$ to $R_O$. We choose MSE as the metric:
\begin{equation}\label{lambda}
  \lambda = \mathop{\arg\min}_{\lambda}(f_{\lambda}(R_S)-R_O)^2
\end{equation}
We should use a model with a small amount of parameters to construct $f_{\lambda}(\cdot)$, so that it is convenient to encode the parameters $\lambda$ into the bitstream to ensure the consistency of encoding and decoding.
For the expression form of $f_{\lambda}(\cdot)$, we have tried linear functions and polynomial functions as shown in Fig. \ref{rm_fig}. From the red box on the left, it can be found that only using the CNN filter (see red dotted line) may lead to a decrease in coding performance, this proves that directly using CNN filters for inter frames may degrade video quality.
And the performance is improved after adopting RM. It is noticeable that there is little difference in performance between different polynomial functions. So we choose simple linear functions to build RM.
\begin{equation}\label{lambda2}
  \lambda = \mathop{\arg\min}_{\lambda}(\lambda R_S-R_O)^2
\end{equation}
So we add $X$ and the output of RM $\hat{R}_S$ to get the filtered frame $\hat{Y}_S$. After sending it to SAO, the entire filtering process is completed.
\begin{equation}\label{rm_output}
  \hat{Y}_S = X + \hat{R}_S = X + \lambda R_S
\end{equation}
%However, the float number of $\lambda$ needs 4 bytes for each frame. To reduce the consumptions of the required bits,
%To find the best parameters, one general solution is to differentiate the loss with respect to $\lambda$.
We quantify the candidate $\lambda$ with $n$ bits for each component, where $\lambda=i/(2^n-1), i \in 0,1,...,2^n-1$. In the implementation, the number of required bits $n$ is set to $5$, so each frame needs 15 bits for the RM module.
And a rate-distortion optimization (RDO) process is designed to find the best $\lambda$. The regular mode of CABAC is used to code $\lambda$. RM does not need specific models for inter frames or additional classifiers for each CTU. What's more, it is independent of the proposed network and can be combined with other learning-based filters to alleviate the over-smoothing problem as well.
%As we known, the frame-level RD cost is the weighted sum of the distortion and the coding bits. The needed coding bits is the same for each frame in our design, so we save the $\lambda$ with least square errors into the bitstreams. The filtered frames  $\widehat{r}(t)$ can be obtained in both encoder and decoder sides.
%\begin{align}\label{mseWiener}
%  MSE &= E[((x(t)+r(t))-\widehat{r}(t))^2]\notag\\
%  &=E[((x(t)+r(t))-(\lambda l(t)+(1-\lambda)r(t)))^2]\notag\\
%  &=E[(x(t)+\lambda(r(t)-l(t)))^2]
%\end{align}
%and,
%\begin{equation}\label{derivative}
%  \frac{\partial MSE}{\partial \lambda} = E[2(r(t)-l(t))(x(t)+\lambda(r(t)-l(t)))]
%\end{equation}
%Set the equation \ref{derivative} equal to zero and we can get the optimal $\lambda$.

\begin{figure}[tbp]
\centering
\subfigure[Serial structure]{
\begin{minipage}[t]{\linewidth}
\centering
\includegraphics[scale=0.5]{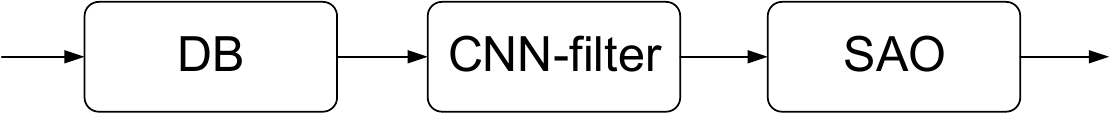}
\end{minipage}%
}%

\subfigure[Parallel structure]{
\begin{minipage}[t]{\linewidth}
\centering
\includegraphics[scale=0.5]{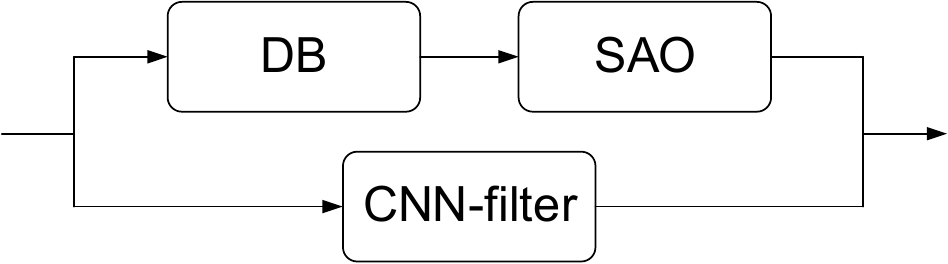}
\end{minipage}%
}%

\subfigure[Proposed structure]{
\begin{minipage}[t]{\linewidth}
\centering
\includegraphics[scale=0.5]{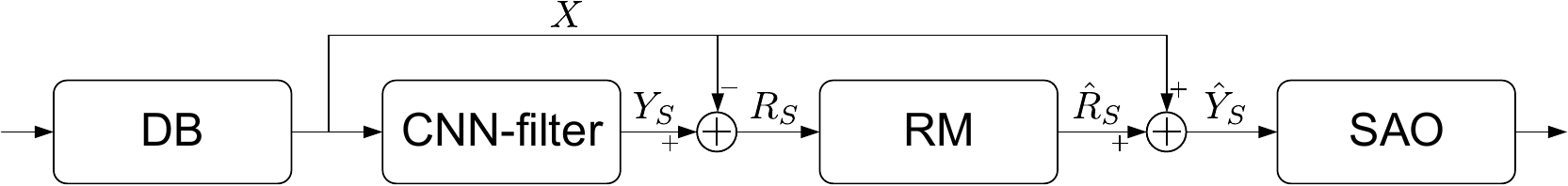}
\end{minipage}%
}%
\centering
\caption{The schemes of the different frameworks with CNN-based filter. }\label{rmfig}
\end{figure}

Different from previous strategy \cite{SDLA} of choosing one between traditional filtering and learning-based filtering, RM uses a serial structure and makes full use of both these two kinds of filtering as shown in Fig. \ref{rmfig}. From the perspective of reconstructed frames, the proposed RM can be interpreted as a post-processing module that fully utilizes the advantages of both distorted reconstruction and learned filtered output. The full use of these two aspects makes RM have excellent performance.
%The $r(t)$ has more details but also some artifacts, while $l(t)$ maybe a little over smooth.
For example, we assume that the reference frame is a frame filtered by a learning-based filter, so if the current frame and the reference frame are almost identical, the current frame does not need to use all the filters.
Conversely, if the current frame and the reference frame are completely different, it is easy to produce artificial imprints because of the distorted residue, so the filters should be used in this case.
For a specific frame, however, it is often difficult to obtain an accurate judgment about whether to use the filters by using its encoded information, such as residuals or motion vectors. Considering the good generalization ability of traditional filters, we keep them working and focused on the CNN filter.
So we introduce a parametric module RM, which uses an RDO process to give an appropriate filtering effect of the CNN filter.
From (\ref{lambda}), it can be observed that the filtering strength varies with the change of the $\lambda$. So we can traverse all of the candidate $\lambda$ and code the one with the smallest reconstruction error into bitstreams. We can also derivative the objective function to obtain the optimal parameters, and code the quantized parameters in the bitstream. In this case, we need to consider the influence of parameter quantization, those mapping functions that are sensitive to quantization noise, such as high-order polynomials, should be abandoned. Otherwise, this may result in larger quantization errors in the decoded frames.

% Almost all of the inter frames can be benefitted from this method and get filtered at a different level.

\section{Experimental Results}

\begin{table}[!tbp]
  \centering
  \caption{Experimental Environment }\label{testcondition}%添加标题 设置标签
  \begin{tabular}{l|l}
    \Xhline{1.0pt}
    Items  & Specification    \bigstrut\\
   \hline \hline
   Optimizer & Adam \cite{kingma2014adam}  \bigstrut\\
   \hline
   Processor &  Intel Xeon Gold 6134 at 3.20 GHz\bigstrut\\
   \hline
   GPU & NVIDIA GeForce RTX 2080 \bigstrut\\
   \hline
   Operating system &  CentOS Linux release 7.6.1810 \bigstrut\\
   \hline
   HM version & 16.16 \bigstrut\\
   \hline
   DNN framework &  Keras 2.2.4 \cite{chollet2015keras} and TensorFlow 1.12.0  \cite{tensorflow} \bigstrut\\

   %DNN framework &  Keras 2.2.4 and TensorFlow 1.12.0 \\

    \Xhline{1.0pt}
    %\bottomrule
  \end{tabular}
\end{table}

\subsection{Experimental Setting}
For the experiment, we mainly focus on objective quality, subjective quality, complexity, and ablation studies to illustrate the performance of our model.
Nine hundred pictures from DIV2K \cite{DIV2K} are cropped into the resolution of $1024\times 1024$, and then down-sampled to $512\times 512$.
These two sets of pictures are spliced into two videos as our training sets.
Only the luminance component is used for training, and the chrominance components are also tested by using the proposed model.
The patch size in training is $32\times 32$ of H.265/HEVC and $64 \times 64$ of H.266/VVC, which is consistent with the largest size of the TU. Considering that the reconstructed images with different QPs often have different degrees of distortion and artifacts, the whole QP band is divided into four parts, below 24, 25 to 29, 30 to 34, and above 35. So four proprietary models are trained for each QP band. The parameter initialization method is normal distribution \cite{he2015delving} for both the teacher model and the proposed model.
The training epochs $n_1$ and $n_3$ are both set to 50.
We use more training epochs for the model with higher QP in the special initialization phase because there are often more artifacts in the reconstructed images with higher QP. Specifically, parameters $n_2$ is set as $10$ for the lower QPs and $20$ for the higher QPs.
After the training phase, we save the trained model and call it to infer in
HEVC reference software (HM) and video coding test model (VTM).
In the test phase, the first 64 frames from HEVC test sequences are used to evaluate the generalization ability of our model. We test four different configurations with default settings, including all-intra (AI), low-delay-B (LDB), low-delay-P (LDP), and random-access (RA) for H.265/HEVC anchor. For the H.266/VVC anchor, we test it with default AI and RA configurations. Four typical QPs in common test conditions are tested, including 22, 27, 32, 37.
The other important test conditions are shown in Table \ref{testcondition}.
For a fair comparison with previous works, we use the coding experimental results from their original papers. The complexities of the reference papers are tested on our local server to avoid the influence of the hardware platforms.

\subsection{Experiment on H.265/HEVC}\label{hmtest}

% Table generated by Excel2LaTeX from sheet 'Sum'
\begin{table*}[tbp]
  \centering
  \caption{BD-rate Reduction of the Proposed Method than HM-16.16 Anchor}
    \begin{tabular}{c|c|r|r|r|r|r|r|r|r|r|r|r|r}
    \Xhline{1.0pt}
    \multicolumn{2}{c|}{\multirow{2}[4]{*}{Sequences}} & \multicolumn{3}{c|}{AI} & \multicolumn{3}{c|}{LDB} & \multicolumn{3}{c|}{LDP} & \multicolumn{3}{c}{RA} \bigstrut\\
\cline{3-14}    \multicolumn{2}{c|}{} & \multicolumn{1}{c|}{Y} & \multicolumn{1}{c|}{U} & \multicolumn{1}{c|}{V} & \multicolumn{1}{c|}{Y} & \multicolumn{1}{c|}{U} & \multicolumn{1}{c|}{V} & \multicolumn{1}{c|}{Y} & \multicolumn{1}{c|}{U} & \multicolumn{1}{c|}{V} & \multicolumn{1}{c|}{Y} & \multicolumn{1}{c|}{U} & \multicolumn{1}{c}{V} \bigstrut\\
    \hline  \hline
    \multirow{2}[4]{*}{ClassA} & \multicolumn{1}{l|}{Traffic} & -7.3\% & -3.4\% & -4.7\% & -4.6\% & -2.4\% & -0.9\% & -4.3\% & -3.4\% & -1.5\% & -6.4\% & -3.6\% & -2.7\% \bigstrut\\
\cline{2-14}          & \multicolumn{1}{l|}{PeopleOnStreet} & -6.8\% & -7.1\% & -6.9\% & -4.5\% & -0.6\% & -0.9\% & -3.1\% & -4.4\% & -2.4\% & -6.1\% & -4.6\% & -5.0\% \bigstrut\\
    \hline
    \multirow{5}[10]{*}{ClassB} & \multicolumn{1}{l|}{Kimono} & -4.9\% & -2.6\% & -2.5\% & -4.6\% & -7.5\% & -4.7\% & -7.3\% & -11.5\% & -6.7\% & -4.2\% & -5.7\% & -3.4\% \bigstrut\\
\cline{2-14}          & \multicolumn{1}{l|}{ParkScene} & -5.5\% & -3.2\% & -2.3\% & -1.9\% & -0.3\% & -0.7\% & -1.5\% & -0.5\% & -0.7\% & -3.8\% & -0.6\% & -0.3\% \bigstrut\\
\cline{2-14}          & \multicolumn{1}{l|}{Cactus} & -5.3\% & -4.1\% & -10.1\% & -4.4\% & -3.7\% & -4.4\% & -5.4\% & -5.6\% & -5.8\% & -6.8\% & -9.5\% & -7.2\% \bigstrut\\
\cline{2-14}          & \multicolumn{1}{l|}{BasketballDrive} & -4.3\% & -8.9\% & -11.7\% & -3.4\% & -4.2\% & -7.8\% & -6.0\% & -9.2\% & -11.7\% & -4.4\% & -4.4\% & -8.9\% \bigstrut\\
\cline{2-14}          & \multicolumn{1}{l|}{BQTerrace} & -3.7\% & -4.3\% & -4.8\% & -6.1\% & -2.1\% & -2.4\% & -10.6\% & -4.5\% & -3.9\% & -8.8\% & -3.8\% & -3.3\% \bigstrut\\
    \hline
    \multirow{4}[8]{*}{ClassC} & \multicolumn{1}{l|}{BasketballDrill} & -8.0\% & -11.7\% & -14.1\% & -2.8\% & -4.9\% & -4.6\% & -3.4\% & -5.6\% & -6.2\% & -4.2\% & -8.0\% & -9.7\% \bigstrut\\
\cline{2-14}          & \multicolumn{1}{l|}{BQMall} & -6.0\% & -6.3\% & -7.2\% & -3.8\% & -3.2\% & -4.7\% & -4.6\% & -4.5\% & -5.6\% & -5.1\% & -4.7\% & -5.1\% \bigstrut\\
\cline{2-14}          & \multicolumn{1}{l|}{PartyScene} & -3.7\% & -4.8\% & -5.7\% & -0.8\% & -0.1\% & -0.2\% & -1.8\% & -0.4\% & -0.4\% & -1.7\% & -1.4\% & -2.0\% \bigstrut\\
\cline{2-14}          & \multicolumn{1}{l|}{RaceHorses} & -3.9\% & -6.9\% & -12.0\% & -4.1\% & -6.6\% & -11.3\% & -4.2\% & -7.9\% & -12.7\% & -4.7\% & -9.7\% & -14.2\% \bigstrut\\
    \hline
    \multirow{4}[8]{*}{ClassD} & \multicolumn{1}{l|}{BasketballPass} & -6.5\% & -7.3\% & -10.3\% & -4.4\% & -3.3\% & -4.6\% & -4.3\% & -4.8\% & -5.8\% & -3.9\% & -4.7\% & -6.1\% \bigstrut\\
\cline{2-14}          & \multicolumn{1}{l|}{BQSquare} & -4.2\% & -3.0\% & -6.8\% & -2.4\% & -1.6\% & -2.8\% & -4.1\% & -1.8\% & -2.9\% & -2.4\% & -1.0\% & -2.9\% \bigstrut\\
\cline{2-14}          & \multicolumn{1}{l|}{BlowingBubbles} & -5.3\% & -9.3\% & -9.8\% & -3.6\% & -5.8\% & -2.1\% & -3.9\% & -5.4\% & -1.8\% & -4.0\% & -6.1\% & -4.4\% \bigstrut\\
\cline{2-14}          & \multicolumn{1}{l|}{RaceHorses} & -7.5\% & -10.5\% & -14.6\% & -6.3\% & -5.2\% & -10.3\% & -6.7\% & -7.4\% & -10.8\% & -6.8\% & -9.4\% & -12.2\% \bigstrut\\
    \hline
    \multirow{6}[12]{*}{ClassE} & \multicolumn{1}{l|}{Vidyo1} & -8.9\% & -8.7\% & -10.5\% & -6.7\% & -9.0\% & -9.6\% & -7.4\% & -9.4\% & -8.9\% & -8.1\% & -8.4\% & -9.7\% \bigstrut\\
\cline{2-14}          & \multicolumn{1}{l|}{Vidyo3} & -7.0\% & -5.2\% & -5.3\% & -4.0\% & -5.9\% & -3.1\% & -4.6\% & -6.3\% & -2.5\% & -6.5\% & -4.1\% & -5.1\% \bigstrut\\
\cline{2-14}          & \multicolumn{1}{l|}{Vidyo4} & -6.3\% & -10.1\% & -10.8\% & -3.8\% & -11.5\% & -10.9\% & -3.9\% & -12.1\% & -11.2\% & -5.6\% & -9.8\% & -10.1\% \bigstrut\\
\cline{2-14}          & \multicolumn{1}{l|}{FourPeople} & -9.4\% & -8.1\% & -9.0\% & -8.6\% & -9.2\% & -9.4\% & -9.0\% & -9.7\% & -10.8\% & -9.4\% & -7.7\% & -8.1\% \bigstrut\\
\cline{2-14}          & \multicolumn{1}{l|}{Johnny} & -8.3\% & -12.3\% & -11.0\% & -7.0\% & -11.4\% & -9.1\% & -9.6\% & -13.1\% & -10.7\% & -8.3\% & -10.9\% & -9.7\% \bigstrut\\
\cline{2-14}          & \multicolumn{1}{l|}{KristenAndSara} & -8.6\% & -10.2\% & -11.1\% & -7.7\% & -8.3\% & -8.6\% & -8.3\% & -10.1\% & -11.2\% & -8.2\% & -8.9\% & -9.6\% \bigstrut\\
    \hline
    \multicolumn{2}{c|}{Average} & \textbf{-6.3\%} & \textbf{-7.0\%} & \textbf{-8.6\%} & \textbf{-4.5\%} & \textbf{-5.1\%} & \textbf{-5.4\%} & \textbf{-5.4\%} & \textbf{-6.6\%} & \textbf{-6.4\%} & \textbf{-5.7\%} & \textbf{-6.1\%} & \textbf{-6.6\%} \bigstrut\\
    \Xhline{1.0pt}
    \end{tabular}%
  \label{results}%
\end{table*}%
\begin{table*}[tbp]
  \centering
  \caption{BD-rate Reduction and Complexity (GPU) of the Proposed Method compared with Previous Works \cite{jia2019content,VRCNN} in AI Configuration}
  \setlength{\tabcolsep}{1.5mm}{
    \begin{tabular}{c|c|c|c|c|c|c|c|c|c|c|c|c|c|c|c}
    \Xhline{1.0pt}
    \multirow{2}[4]{*}{Sequences} & \multicolumn{5}{c|}{Jia et al. \cite{jia2019content}} & \multicolumn{5}{c|}{VR-CNN \cite{VRCNN}} & \multicolumn{5}{c}{Proposed model} \bigstrut\\
\cline{2-16}      & Y & U & V & $\Delta T_{enc}$ & $\Delta T_{dec}$ & Y & U & V & $\Delta T_{enc}$ & $\Delta T_{dec}$ & Y & U & V & $\Delta T_{enc}$ & $\Delta T_{dec}$ \bigstrut\\
    \hline \hline
    ClassA & \multicolumn{1}{r|}{-4.7\%} & \multicolumn{1}{r|}{-3.3\%} & \multicolumn{1}{r|}{-2.6\%} & \multicolumn{1}{r|}{108.1\%} & \multicolumn{1}{r|}{734.9\%} & \multicolumn{1}{r|}{-5.5\%} & \multicolumn{1}{r|}{-4.7\%} & \multicolumn{1}{r|}{-4.9\%} & \multicolumn{1}{r|}{108.3\%} & \multicolumn{1}{r|}{561.1\%} & \multicolumn{1}{r|}{-7.1\%} & \multicolumn{1}{r|}{-5.4\%} & \multicolumn{1}{r|}{-5.9\%} & \multicolumn{1}{r|}{105.8\%} & \multicolumn{1}{r}{281.0\%} \bigstrut\\
    \hline
    ClassB & \multicolumn{1}{r|}{-3.5\%} & \multicolumn{1}{r|}{-2.8\%} & \multicolumn{1}{r|}{-3.0\%} & \multicolumn{1}{r|}{109.0\%} & \multicolumn{1}{r|}{659.8\%} & \multicolumn{1}{r|}{-3.3\%} & \multicolumn{1}{r|}{-3.2\%} & \multicolumn{1}{r|}{-3.7\%} & \multicolumn{1}{r|}{110.3\%} & \multicolumn{1}{r|}{505.3\%} & \multicolumn{1}{r|}{-4.8\%} & \multicolumn{1}{r|}{-4.8\%} & \multicolumn{1}{r|}{-6.4\%} & \multicolumn{1}{r|}{106.2\%} & \multicolumn{1}{r}{265.2\%} \bigstrut\\
    \hline
    ClassC & \multicolumn{1}{r|}{-3.4\%} & \multicolumn{1}{r|}{-3.5\%} & \multicolumn{1}{r|}{-5.0\%} & \multicolumn{1}{r|}{113.1\%} & \multicolumn{1}{r|}{894.9\%} & \multicolumn{1}{r|}{-5.0\%} & \multicolumn{1}{r|}{-5.5\%} & \multicolumn{1}{r|}{-6.9\%} & \multicolumn{1}{r|}{113.0\%} & \multicolumn{1}{r|}{685.1\%} & \multicolumn{1}{r|}{-5.4\%} & \multicolumn{1}{r|}{-7.5\%} & \multicolumn{1}{r|}{-9.9\%} & \multicolumn{1}{r|}{106.5\%} & \multicolumn{1}{r}{326.3\%} \bigstrut\\
    \hline
    ClassD & \multicolumn{1}{r|}{-3.2\%} & \multicolumn{1}{r|}{-4.7\%} & \multicolumn{1}{r|}{-6.0\%} & \multicolumn{1}{r|}{128.9\%} & \multicolumn{1}{r|}{1406.0\%} & \multicolumn{1}{r|}{-5.4\%} & \multicolumn{1}{r|}{-6.4\%} & \multicolumn{1}{r|}{-8.1\%} & \multicolumn{1}{r|}{121.6\%} & \multicolumn{1}{r|}{1047.1\%} & \multicolumn{1}{r|}{-5.9\%} & \multicolumn{1}{r|}{-7.8\%} & \multicolumn{1}{r|}{-10.5\%} & \multicolumn{1}{r|}{114.4\%} & \multicolumn{1}{r}{548.0\%} \bigstrut\\
    \hline
    ClassE & \multicolumn{1}{r|}{-5.8\%} & \multicolumn{1}{r|}{-4.1\%} & \multicolumn{1}{r|}{-5.2\%} & \multicolumn{1}{r|}{112.3\%} & \multicolumn{1}{r|}{1110.2\%} & \multicolumn{1}{r|}{-6.5\%} & \multicolumn{1}{r|}{-5.5\%} & \multicolumn{1}{r|}{-5.6\%} & \multicolumn{1}{r|}{111.1\%} & \multicolumn{1}{r|}{836.7\%} & \multicolumn{1}{r|}{-8.1\%} & \multicolumn{1}{r|}{-9.2\%} & \multicolumn{1}{r|}{-9.7\%} & \multicolumn{1}{r|}{107.2\%} & \multicolumn{1}{r}{401.1\%} \bigstrut\\
    \hline
    Average & \multicolumn{1}{r|}{-4.1\%} & \multicolumn{1}{r|}{-3.7\%} & \multicolumn{1}{r|}{-4.4\%} & \multicolumn{1}{r|}{114.3\%} & \multicolumn{1}{r|}{961.2\%} & \multicolumn{1}{r|}{-5.1\%} & \multicolumn{1}{r|}{-5.1\%} & \multicolumn{1}{r|}{-5.8\%} & \multicolumn{1}{r|}{112.9\%} & \multicolumn{1}{r|}{727.0\%} & \multicolumn{1}{r|}{\textbf{-6.3\%}} & \multicolumn{1}{r|}{\textbf{-7.0\%}} & \multicolumn{1}{r|}{\textbf{-8.6\%}} & \multicolumn{1}{r|}{\textbf{108.0\%}} & \multicolumn{1}{r}{\textbf{364.3\%}} \bigstrut\\
    \hline
    FLOPs & \multicolumn{5}{c|}{334.84G } & \multicolumn{5}{c|}{50.39G } & \multicolumn{5}{c}{\textbf{10.51G}} \bigstrut\\
    \hline
    Parameters & \multicolumn{5}{c|}{362,753} & \multicolumn{5}{c|}{54,512} & \multicolumn{5}{c}{\textbf{11,114}} \bigstrut\\
    \hline
    Model size & \multicolumn{5}{c|}{1.38MB } & \multicolumn{5}{c|}{220KB } & \multicolumn{5}{c}{\textbf{58KB}} \bigstrut\\
    \Xhline{1.0pt}
    \end{tabular}%
    }
  \label{comparsion}%
\end{table*}%

% Table generated by Excel2LaTeX from sheet 'Sum'
\begin{table*}[tbp]
  \centering
  \caption{Overall BD-rate Comparison of Previous Methods\cite{jia2019content,IFCNN,STResNet} in LDB, LDP, and RA Configuration}
    \begin{tabular}{rrrrrrrrrr}
    \Xhline{1.0pt}
    \multicolumn{1}{c|}{\multirow{2}[4]{*}{Methods}} & \multicolumn{3}{c|}{LDB} & \multicolumn{3}{c|}{LDP} & \multicolumn{3}{c}{RA} \bigstrut\\
\cline{2-10}    \multicolumn{1}{c|}{} & \multicolumn{1}{c|}{Y} & \multicolumn{1}{c|}{U} & \multicolumn{1}{c|}{V} & \multicolumn{1}{c|}{Y} & \multicolumn{1}{c|}{U} & \multicolumn{1}{c|}{V} & \multicolumn{1}{c|}{Y} & \multicolumn{1}{c|}{U} & \multicolumn{1}{c}{V} \bigstrut\\
    \hline  \hline
    \multicolumn{1}{c|}{Jia et al. \cite{jia2019content}} & \multicolumn{1}{r|}{\textbf{-6.0\%}} & \multicolumn{1}{r|}{-2.9\%} & \multicolumn{1}{r|}{-3.5\%} & \multicolumn{1}{r|}{-4.7\%} & \multicolumn{1}{r|}{-1.0\%} & \multicolumn{1}{r|}{-1.2\%} & \multicolumn{1}{r|}{\textbf{-6.0\%}} & \multicolumn{1}{r|}{-3.2\%} & -3.8\% \bigstrut\\
    \hline
    \multicolumn{1}{c|}{Our network + RM} & \multicolumn{1}{r|}{-4.5\%} & \multicolumn{1}{r|}{\textbf{-5.1\%}} & \multicolumn{1}{r|}{\textbf{-5.4\%}} & \multicolumn{1}{r|}{\textbf{-5.4\%}} & \multicolumn{1}{r|}{\textbf{-6.6\%}} & \multicolumn{1}{r|}{\textbf{-6.4\%}} & \multicolumn{1}{r|}{-5.7\%} & \multicolumn{1}{r|}{\textbf{-6.1\%}} & \textbf{-6.6\%} \bigstrut\\
    \hline
    \multicolumn{1}{c|}{Our network + Frame control\cite{IFCNN} } & \multicolumn{1}{r|}{-3.7\%} & \multicolumn{1}{r|}{-3.3\%} & \multicolumn{1}{r|}{-3.2\%} & \multicolumn{1}{r|}{-4.4\%} & \multicolumn{1}{r|}{-4.6\%} & \multicolumn{1}{r|}{-3.9\%} & \multicolumn{1}{r|}{-4.6\%} & \multicolumn{1}{r|}{-4.8\%} & -5.1\% \bigstrut\\
    \hline
    \multicolumn{1}{c|}{Our network + CTU control \cite{STResNet}} & \multicolumn{1}{r|}{-4.1\%} & \multicolumn{1}{r|}{-4.4\%} & \multicolumn{1}{r|}{-4.9\%} & \multicolumn{1}{r|}{-4.6\%} & \multicolumn{1}{r|}{-5.8\%} & \multicolumn{1}{r|}{-5.9\%} & \multicolumn{1}{r|}{-4.5\%} & \multicolumn{1}{r|}{-5.1\%} & -5.8\% \bigstrut\\

    \Xhline{1.0pt}
    \end{tabular}%
  \label{pbFilter}%
\end{table*}%

\subsubsection{Objective Evaluation}
%In our experiment, DB and SAO are disabled because the proposed model is utilized to replace the whole in-loop filtering.
In this subsection, the objective evaluation is conducted to evaluate the performance of our proposed model.
The experimental results compared with the HM-16.16 anchor are shown in Table \ref{results}.
For the luminance component, the proposed model achieves 6.3\%, 4.5\%, 5.4\%, and 5.7\% BD-rate reduction compared with HEVC baseline under AI, LDB, LDP, and RA configuration, respectively. For chrominance components, the proposed model achieves more BD-rate reduction than the luminance component. It demonstrates the generalization ability for the proposed model because we only use the luminance components of intra samples for training.
Furthermore, the comparisons with the previous works\cite{jia2019content, VRCNN} are conducted and the BD-rate reduction is shown in Table \ref{comparsion}. It can be seen that our model achieves more BD-rate reduction for AI configuration.
%Specific to individual sequences, the student model almost achieve better BD-rate reduction for all video sequences and components.% At the same time, for Kimono and BQTerrace, the performance of the student model and VR-CNN are not as good as other sequences. This is because the texture of these two sequences is complicated, and for those pictures with more complex textures, the filtering will lead to the loss of high-frequency components.
%The performance comparison of CTU-level \cite{dai2018cnn} and the proposed merge algorithm are shown in Table \ref{pbFilter}.

% We believe that the performance of the learning-based filter is mainly reflected in the intra frames. On the one hand, for those inter blocks with obvious artifacts, it makes almost no difference to filter them compared with intra frames (For example, the intra block in inter frame). On the other hand, the inter blocks with few artifacts themselves have a relatively high quality and do not have to be filtered.

For the performance evaluation of inter configurations, we introduce the comparison of our proposed model with frame-level control \cite{IFCNN}, CTU-level control \cite{STResNet}, and Jia \textit{et al.} \cite{jia2019content} as shown in Table \ref{pbFilter}.
To compare fairly, we select the same padding and use the proposed model to test the different control methods.
From the experiment results,
it can be seen that our proposed model achieves about 1\% extra BD-rate reduction than both CTU-level and frame-level control for all inter configurations. Compared with Jia \textit{et al.} \cite{jia2019content}, our model achieves comparative BD-rate reduction in inter configurations. For the chrominance components, our model achieves about 3\% extra BD-rate reduction, it further demonstrates the generalization ability of our model.

\begin{figure*}[tbp]
\centering
\subfigure[HM Rec.]{
\begin{minipage}[t]{0.325\linewidth}
\centering
\includegraphics[width=5.6cm]{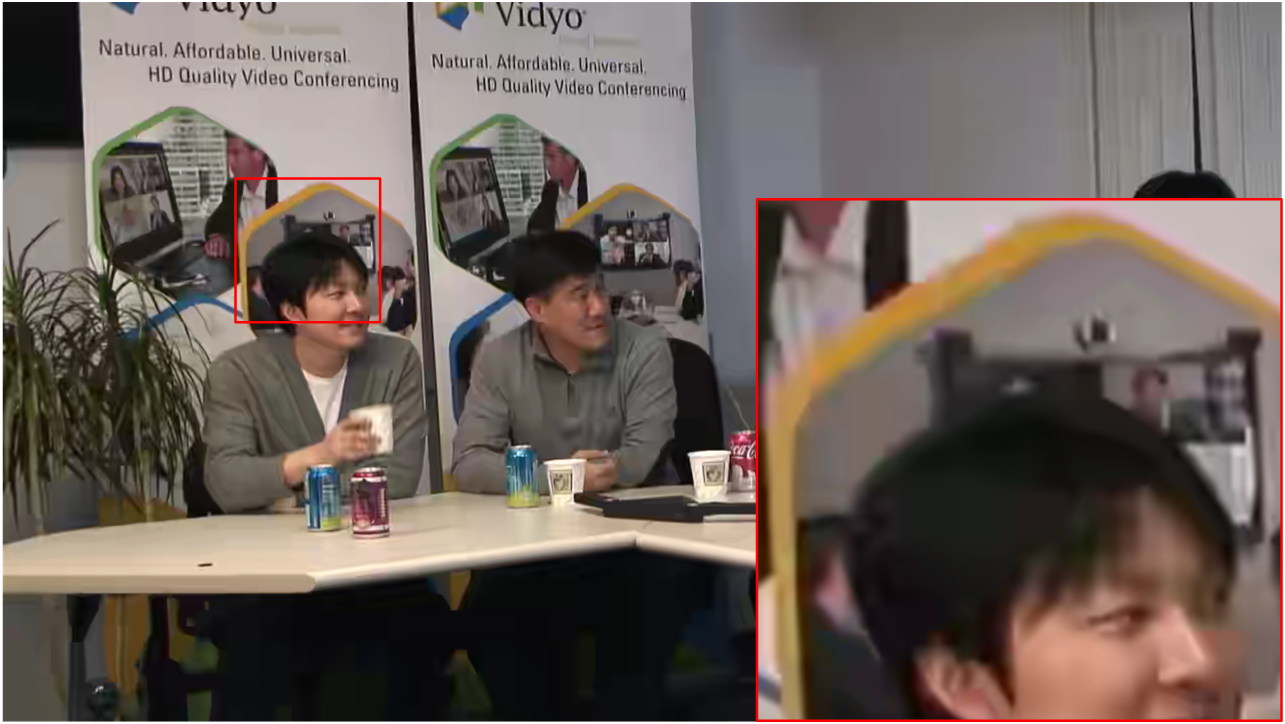}
\end{minipage}%
}%
\subfigure[Jia \textit{et al.} \cite{jia2019content}]{
\begin{minipage}[t]{0.325\linewidth}
\centering
\includegraphics[width=5.6cm]{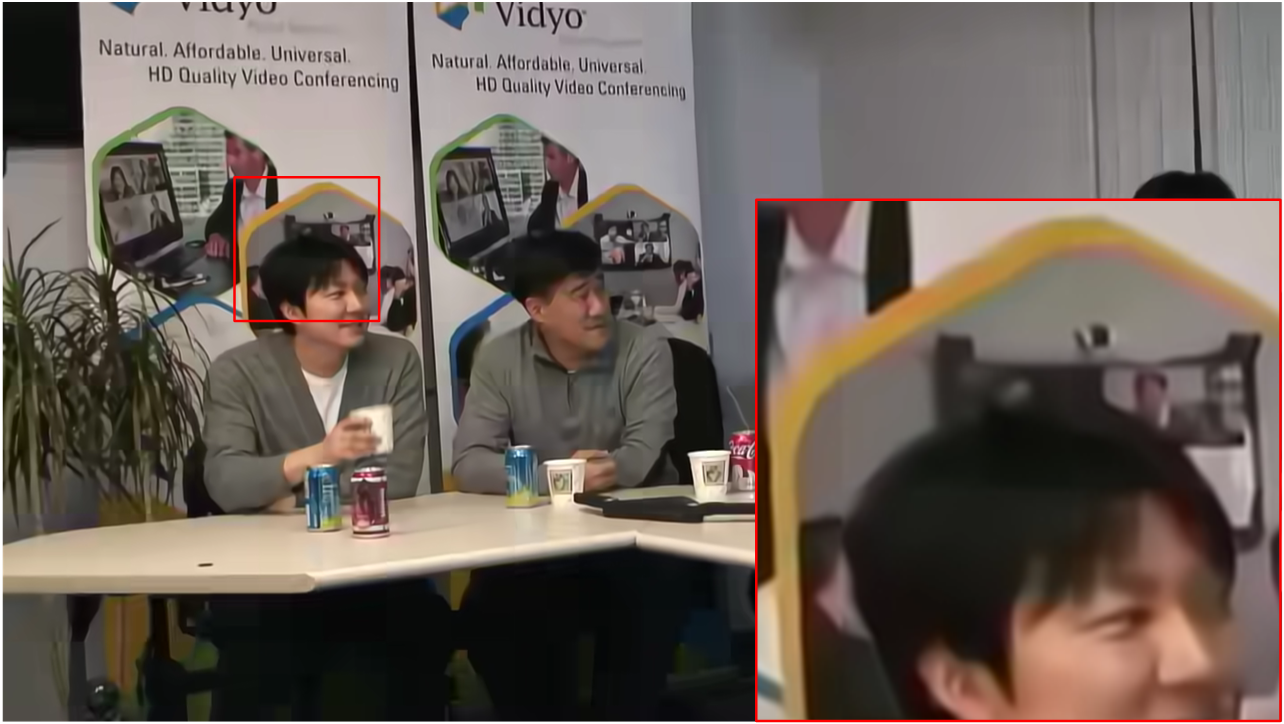}
\end{minipage}%
}%
\subfigure[Proposed model]{
\begin{minipage}[t]{0.325\linewidth}
\centering
\includegraphics[width=5.6cm]{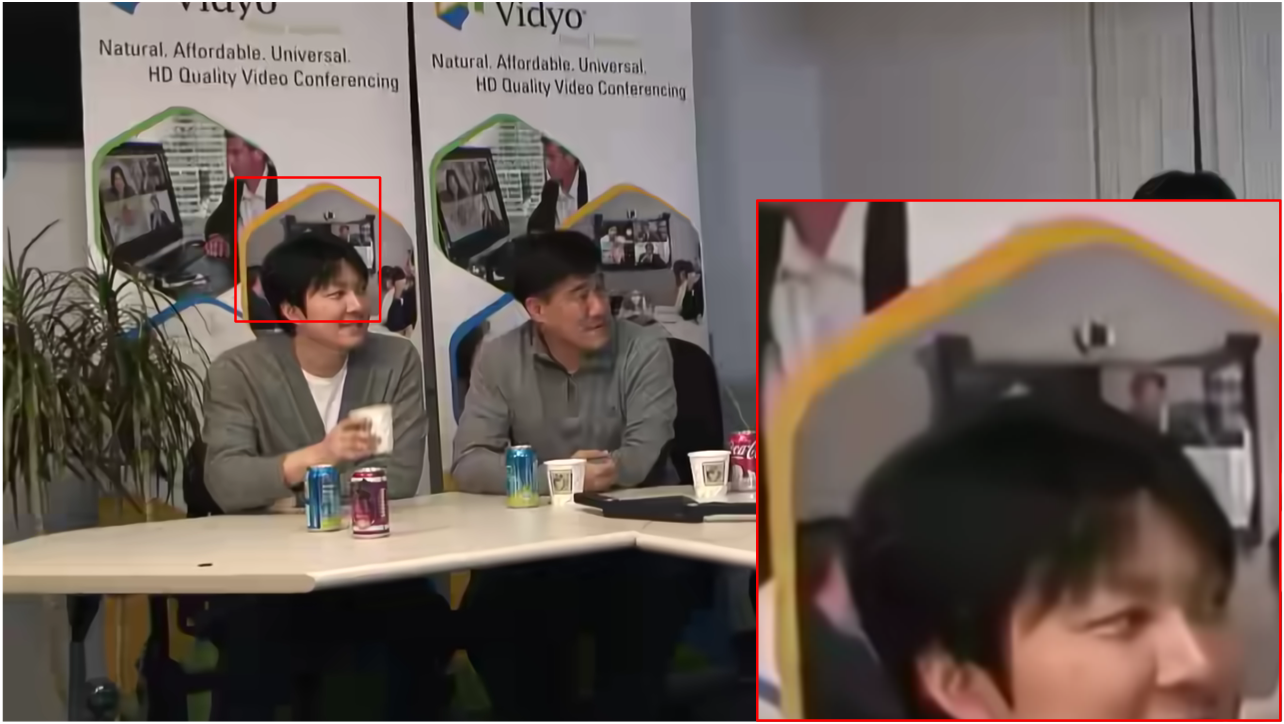}
\end{minipage}%
}%
\centering
\caption{Visual quality comparison of Jia \textit{et al.} \cite{jia2019content} and the proposed model for AI configuration. The test qp is 37 and this is the 1st frame for FourPeople(Anchor HM-16.9). }\label{subjective1}
\end{figure*}

\begin{figure*}[tbp]
\centering
\subfigure[HM Rec.]{
\begin{minipage}[t]{0.325\linewidth}
\centering
\includegraphics[width=5.6cm]{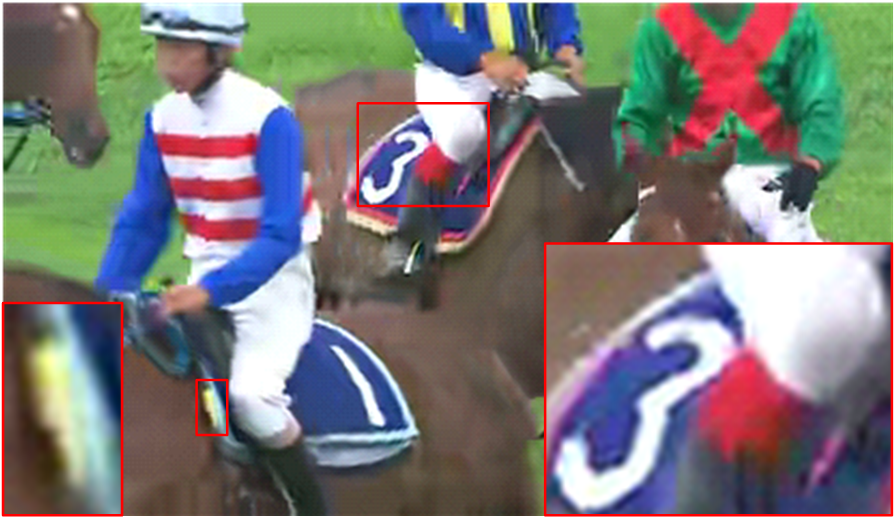}
\end{minipage}%
}%
\subfigure[CTU-level control \cite{dai2018cnn}]{
\begin{minipage}[t]{0.325\linewidth}
\centering
\includegraphics[width=5.6cm]{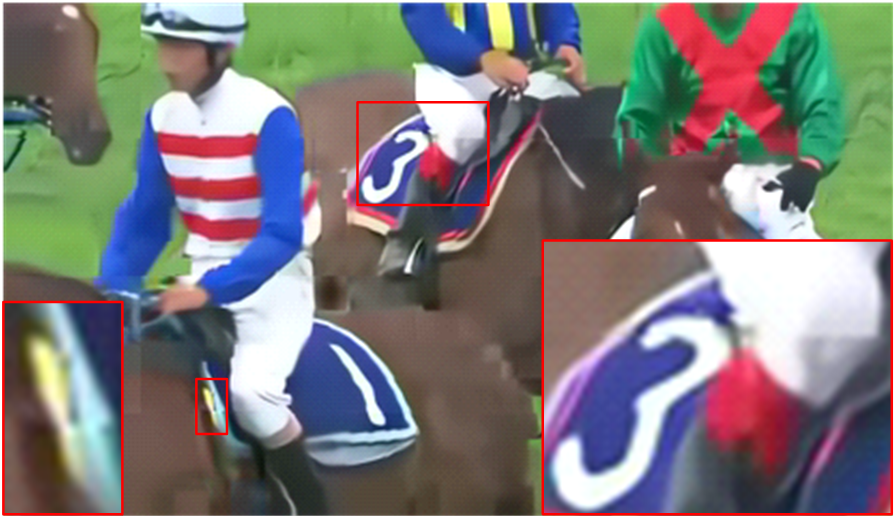}
\end{minipage}%
}%
\subfigure[Proposed model]{
\begin{minipage}[t]{0.325\linewidth}
\centering
\includegraphics[width=5.6cm]{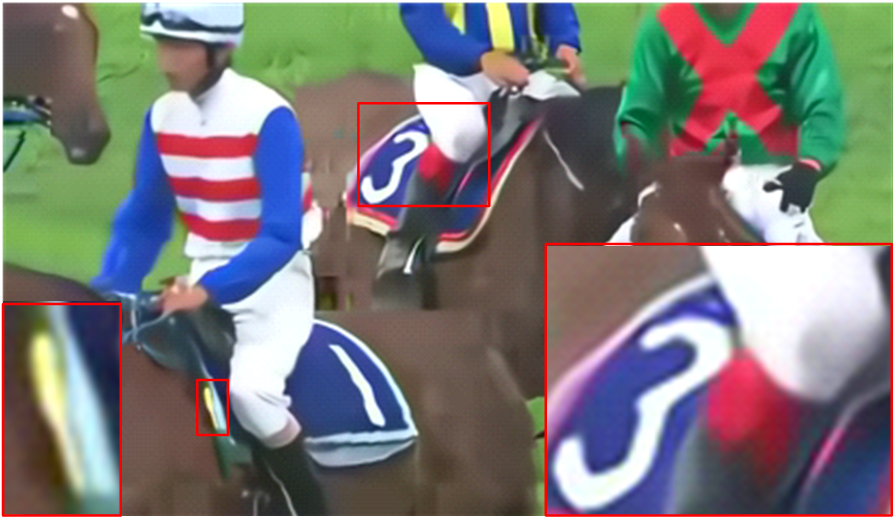}
\end{minipage}%
}%
\centering
\caption{Visual quality comparison of CTU-level control \cite{dai2018cnn} and the proposed model for RA configuration. The test QP is 37 and this is the 16th frame for RaceHorse(Anchor HM-16.16).}\label{subjective2}
\end{figure*}

\subsubsection{Subjective Evaluation}
We also conduct the subjective evaluation as shown in Fig. \ref{subjective1} and Fig. \ref{subjective2}. It can be seen from the experimental results that our model has a great de-artifacts capability.
First, we re-deploy the proposed model in HM-16.9 for a fair subjective evaluation with Jia \textit{et al.} \cite{jia2019content}. From Fig. \ref{subjective1}, it can be found that the various kinds of artifacts in (a) are eliminated by the proposed model and the man's face looks smoother and plump. At the same time, some vertical blocky effects are produced by Jia \textit{et al.} \cite{jia2019content}, probably because it uses different filters for consecutive CTUs while our proposed model uses the same filter for the whole images and have no additional boundaries. Besides, the man's eyes seem to be blurred by \cite{jia2019content} and lead to the degradation of visual quality.
Second, the subjective evaluation for the inter frames is conducted in Fig. \ref{subjective2}.
%Similarly, we use the same test condition in objective evaluation for this test.
The default HM and HM with CTU-level control \cite{dai2018cnn} are used as the anchors.
As shown in Fig. \ref{subjective2}, the contouring and blocky artifacts on the number are eliminated by the proposed model.
For CTU-level control \cite{dai2018cnn} based filtering, the subjective quality of this frame is reduced due to the artificial boundaries on the knee, whereas our proposed model has no boundaries on it and achieves a better visual quality.
To sum up, because our proposed method makes full use of the frame-level filtering strategy, the proposed method has significantly better visual effects than previous CTU-based methods.

\begin{table*}[tbp]
  \centering
  \caption{BD-rate Reduction and Computational Complexity (GPU) of the Proposed Method than VTM-6.3 Anchor}
    \begin{tabular}{c|c|r|r|r|r|r|r|r|r|r|r}
    \Xhline{1.0pt}
    \multicolumn{2}{c|}{\multirow{2}[4]{*}{Sequences}} & \multicolumn{5}{c|}{AI} & \multicolumn{5}{c}{RA} \bigstrut\\
\cline{3-12}    \multicolumn{2}{c|}{} & \multicolumn{1}{c|}{Y} & \multicolumn{1}{c|}{U} & \multicolumn{1}{c|}{V} & \multicolumn{1}{c|}{$\Delta T_{enc}$} & \multicolumn{1}{c|}{$\Delta T_{dec}$} & \multicolumn{1}{c|}{Y} & \multicolumn{1}{c|}{U} & \multicolumn{1}{c|}{V} & \multicolumn{1}{c|}{$\Delta T_{enc}$} & \multicolumn{1}{c}{$\Delta T_{dec}$} \bigstrut\\
    \hline \hline
    \multirow{2}[4]{*}{ClassA} & \multicolumn{1}{l|}{Traffic} & -1.6\% & -0.2\% & -0.4\% & 100.7\% & 234.1\% & -1.1\% & -0.7\% & -0.5\% & 99.6\% & 357.0\% \bigstrut\\
\cline{2-12}      & \multicolumn{1}{l|}{PeopleOnStreet} & -1.3\% & -0.4\% & -0.3\% & 98.0\% & 225.4\% & -0.9\% & -0.1\% & -0.2\% & 99.7\% & 266.3\% \bigstrut\\
    \hline
    \multirow{5}[10]{*}{ClassB} & \multicolumn{1}{l|}{Kimono} & -0.3\% & 0.1\% & -0.3\% & 104.9\% & 317.5\% & -0.2\% & 0.0\% & -0.4\% & 99.8\% & 319.3\% \bigstrut\\
\cline{2-12}      & \multicolumn{1}{l|}{ParkScene} & -1.9\% & 0.1\% & -0.1\% & 108.3\% & 232.4\% & -1.4\% & 0.3\% & -0.2\% & 101.2\% & 302.1\% \bigstrut\\
\cline{2-12}      & \multicolumn{1}{l|}{Cactus} & -1.3\% & -0.5\% & -0.8\% & 100.4\% & 244.5\% & -1.4\% & -1.8\% & -1.7\% & 103.1\% & 343.9\% \bigstrut\\
\cline{2-12}      & \multicolumn{1}{l|}{BasketballDrive} & -0.3\% & -0.8\% & -1.0\% & 103.7\% & 282.7\% & -0.4\% & -0.8\% & -0.5\% & 100.5\% & 321.0\% \bigstrut\\
\cline{2-12}      & \multicolumn{1}{l|}{BQTerrace} & -1.0\% & -0.6\% & -0.6\% & 101.6\% & 228.4\% & -1.9\% & -1.6\% & -1.5\% & 101.6\% & 313.0\% \bigstrut\\
    \hline
    \multirow{4}[8]{*}{ClassC} & \multicolumn{1}{l|}{BasketballDrill} & -2.7\% & -3.8\% & -5.5\% & 101.6\% & 219.2\% & -1.6\% & -3.4\% & -2.9\% & 102.9\% & 270.8\% \bigstrut\\
\cline{2-12}      & \multicolumn{1}{l|}{BQMall} & -2.2\% & -0.8\% & -0.7\% & 100.4\% & 220.6\% & -2.0\% & -1.1\% & -0.5\% & 104.6\% & 285.1\% \bigstrut\\
\cline{2-12}      & \multicolumn{1}{l|}{PartyScene} & -1.8\% & -1.1\% & -1.5\% & 100.5\% & 198.7\% & -1.3\% & -1.6\% & -1.8\% & 101.7\% & 242.7\% \bigstrut\\
\cline{2-12}      & \multicolumn{1}{l|}{RaceHorses} & -0.9\% & -1.1\% & -2.3\% & 101.6\% & 233.5\% & -1.1\% & -1.5\% & -2.5\% & 99.5\% & 243.8\% \bigstrut\\
    \hline
    \multirow{4}[8]{*}{ClassD} & \multicolumn{1}{l|}{BasketballPass} & -2.1\% & -1.4\% & -4.7\% & 99.4\% & 407.5\% & -1.2\% & -2.4\% & -1.0\% & 98.0\% & 406.1\% \bigstrut\\
\cline{2-12}      & \multicolumn{1}{l|}{BQSquare} & -3.0\% & -0.2\% & -1.0\% & 103.0\% & 319.1\% & -3.6\% & -1.0\% & -1.6\% & 105.6\% & 421.0\% \bigstrut\\
\cline{2-12}      & \multicolumn{1}{l|}{BlowingBubbles} & -2.1\% & -1.4\% & -1.0\% & 101.1\% & 352.0\% & -1.6\% & -2.3\% & -2.4\% & 101.0\% & 371.1\% \bigstrut\\
\cline{2-12}      & \multicolumn{1}{l|}{RaceHorses} & -2.8\% & -2.7\% & -4.6\% & 99.6\% & 366.6\% & -2.4\% & -3.1\% & -6.5\% & 98.6\% & 312.1\% \bigstrut\\
    \hline
    \multirow{6}[12]{*}{ClassE} & \multicolumn{1}{l|}{Vidyo1} & -1.3\% & -0.1\% & -0.3\% & 101.5\% & 340.3\% & -1.0\% & -0.1\% & 0.4\% & 102.6\% & 486.6\% \bigstrut\\
\cline{2-12}      & \multicolumn{1}{l|}{Vidyo3} & -1.1\% & 0.2\% & -0.2\% & 101.8\% & 298.8\% & -1.2\% & 1.5\% & 0.9\% & 102.0\% & 457.4\% \bigstrut\\
\cline{2-12}      & \multicolumn{1}{l|}{Vidyo4} & -0.8\% & -0.3\% & -0.2\% & 106.4\% & 291.0\% & -1.0\% & 0.3\% & -1.4\% & 101.3\% & 425.5\% \bigstrut\\
\cline{2-12}      & \multicolumn{1}{l|}{FourPeople} & -2.1\% & -0.5\% & -0.5\% & 99.2\% & 263.3\% & -1.8\% & -0.8\% & -0.7\% & 106.2\% & 449.3\% \bigstrut\\
\cline{2-12}      & \multicolumn{1}{l|}{Johnny} & -1.3\% & -0.4\% & -0.6\% & 99.9\% & 317.9\% & -2.6\% & -1.2\% & -1.1\% & 100.1\% & 452.4\% \bigstrut\\
\cline{2-12}      & \multicolumn{1}{l|}{KristenAndSara} & -1.7\% & -0.6\% & -0.7\% & 101.1\% & 344.5\% & -1.6\% & -0.4\% & -1.4\% & 100.2\% & 428.0\% \bigstrut\\
    \hline
    \multicolumn{2}{c|}{Average} & \textbf{-1.6\%} & \textbf{-0.8\%} & \textbf{-1.3\%} & \textbf{101.6\%} & \textbf{282.8\%} & \textbf{-1.5\%} & \textbf{-1.0\%} & \textbf{-1.3\%} & \textbf{101.4\%} & \textbf{355.9\%} \bigstrut\\
    \Xhline{1.0pt}
    \end{tabular}%
  \label{vtmtest}%
\end{table*}%

\subsubsection{Complexity Analysis} %Our student model is quite light compared to VR-CNN.
%Since the size of videos in class A is too large to be placed in our GPU at one time, only classes B{-}E are tested in this subsection.
As shown in Table \ref{comparsion}, we compare the complexity of Jia \textit{et al.} \cite{jia2019content}, VR-CNN \cite{VRCNN}, and our proposed model from two aspects, including computational complexity and storage consumption.
%the amount of complexity relates to the number of model parameters and the size of input images.
Firstly, for the coding complexity evaluation, we use the following equation to calculate the $\Delta T$:
\begin{equation}\label{delta}
  \Delta T=\frac{T'}{T}
\end{equation}
where $T'$ and $T$ denote the HM coding time with and without the learning-based filter, respectively.
FLOPs in Table \ref{comparsion} are also tested for the frame with a resolution of 720p. Compared with VR-CNN \cite{VRCNN}, the FLOPs of our model is reduced by 79.1\%.
The decoding complexity is reduced by approximately 50\% and the encoding complexity is reduced by 4\%.
The processing time of the proposed model is almost the same for both encoder and decoder. The difference in relative time is caused by that the network inference time accounts for a small proportion of the encoding complexity but comparative for the decoding.

In terms of storage consumption, compared with \cite{VRCNN}, the number of trainable parameters in the proposed model is reduced by 79.6\%. It is almost the same with the reduction of model size because we use the same precision (float32) to save the models. The main reason why our model has relatively fewer parameters is that the design of the proposed model focuses more on complexity instead of performance. For example, we use the DSC as the backbone of the proposed model, whereas previous works \cite{VRCNN,jia2019content} utilize the standard convolution. Meanwhile, we also use many useful methods to limit the model size while maintaining the performance, including BN merge and special initialization of parameters. What's more, our proposed model only needs one learning-based network for both intra and inter frames. So there is no need for additional models in practical applications. Compared with previous works that need multiple models or classifiers, our proposed method reduces the required storage consumption effectively benefit from the RM module.

\subsection{Experiment on H.266/VVC}
To further evaluate the performance of our proposed model, we use the same test condition to test its performance in VTM-6.3. The only difference is that we use the entire DIV2k instead of the down-sampled dataset to train the proposed model.
From the experimental results shown in Table \ref{vtmtest}, it can be found that our model achieves about 1.6\% and 1.5\% BD-rate reduction on the luminance component for AI and RA configurations. For chrominance components, it also achieved similar performance on BD-rate reduction.
In terms of complexity, the proposed method introduces a negligible increase on the encoding side and brings about 3 times complexity to the decoding side.
%Compared with HM, its relative increased complexity is significantly lower.

\begin{table}[tbp]
  \centering
  \caption{Ablation Study of RM (AI, VTM-6.3)}
    \begin{tabular}{c|r|r|r|r|r|r}
    \Xhline{1.0pt}
    \multirow{2}[4]{*}{Sequences} & \multicolumn{3}{c|}{Our network} & \multicolumn{3}{c}{Our network+RM} \bigstrut\\
\cline{2-7}      & \multicolumn{1}{c|}{Y} & \multicolumn{1}{c|}{U} & \multicolumn{1}{c|}{V} & \multicolumn{1}{c|}{Y} & \multicolumn{1}{c|}{U} & \multicolumn{1}{c}{V} \bigstrut\\
    \hline \hline
    ClassA & -0.5\% & -0.1\% & 0.4\% & -1.5\% & -0.3\% & -0.4\% \bigstrut\\
    \hline
    ClassB & 0.3\% & 1.3\% & 0.1\% & -1.0\% & -0.3\% & -0.5\% \bigstrut\\
    \hline
    ClassC & -1.6\% & -1.8\% & -2.8\% & -1.9\% & -1.7\% & -2.5\% \bigstrut\\
    \hline
    ClassD & -2.6\% & -2.2\% & -3.6\% & -2.5\% & -1.4\% & -2.8\% \bigstrut\\
    \hline
    ClassE & -0.3\% & 2.6\% & 1.5\% & -1.4\% & -0.3\% & -0.4\% \bigstrut\\
    \hline
    Average & -0.9\% & 0.3\% & -0.7\% & \textbf{-1.6\%} & \textbf{-0.8\%} & \textbf{-1.3\%} \bigstrut\\
    \Xhline{1.0pt}
    \end{tabular}%
  \label{ablation1}%
\end{table}%

\begin{table}[tbp]
  \centering
  \caption{Ablation Study of Parameter Initialization (AI, VTM-6.3)}\label{ablation2}
  \begin{tabular}{c|c|c|c}
    \Xhline{1.0pt}
   \multirow{2}[4]{*}{Methods} &  \multicolumn{3}{c}{$\Delta$PSNR(dB)} \bigstrut\\
   \cline{2-4}
   & Y  &U&V  \bigstrut \\
   \hline \hline
   Student &0.310&0.231&0.295 \bigstrut \\
   \hline
   Student + MMD \cite{huang2017like}&0.320&0.245&0.313 \bigstrut \\
   \hline
   Student + AT \cite{zagoruyko2016paying}&   \textbf{0.329}&\textbf{0.256}&\textbf{0.328} \bigstrut \\
    \Xhline{1.0pt}
    %\bottomrule
  \end{tabular}
\end{table}

\subsection{Ablation Study} \label{rm_intra_ab}
\subsubsection{RM for intra frames}
RM can effectively improve the generalization ability of learning-based filters. The experiments of RM about the inter frames have been carried out in Section \ref{hmtest}.  Based on VTM here, we further conduct ablation experiments on intra frames to illustrate the performance of RM. Its test setting is the same as before.
From the experiment shown in Table \ref{ablation1}, we can find about 0.8\% BD-rate reduction has been achieved by the RM module. Regarding the performance of class-B, only using the proposed CNN-filter may even have a negative effect and leads to 0.3\% BD-rate increment. But its performance has been well improved after using RM. For most other classes, the performance has also been improved more or less after using RM.

\subsubsection{The initialization of parameters} The 1-st frame of all HEVC test sequences is tested and the overall PSNR increments are shown in Table \ref{ablation2}, where the student model without transfer learning is indicated as "Student" row. MMD and AT in Table \ref{ablation2} represent different transfer learning ways that act on the student model.
By comparing the "Student" row with the other rows, we can find that the PSNR of the student model is improved by both MMD and AT. What's more, the improvements of the chrominance components are more obvious than that of the luminance component.
%As shown in Fig. \ref{loss_fig}, we also plot the training loss and testing loss of the models mentioned above. Although the testing loss has some small variations, it can be seen that both MMD and AT help the student model converge faster. And the student model with AT performs the best in our test.
%What's more, Fig. \ref{loss_fig} proves the generalization ability of the proposed model because the training set and the testing set have the same downtrend although they are different.
%In summary, the student model converges faster and gets extra performance improvement by using knowledge transfer.

\section{Conclusion}
In this paper, a CNN-based low complexity filter is proposed for video coding. The lightweight DSC merged with the batch normalization is used as the backbone.
Based on the transfer learning, attention transfer is utilized to initialize the parameters of the proposed network. By adding a novel parametric module RM after the CNN filter, the generality of the CNN filter is improved and can also handle the filtering problem of inter frames.
What's more, RM is independent of the proposed network and can also combine with other learning-based filters to alleviate the over-smoothing problem.
The experimental results show our proposed model achieves excellent performance in terms of both BD-rate and complexity. For HEVC test sequences, our proposed model achieves about 1.2\% BD-rate reduction and 79.1\% FLOPs than VR-CNN anchor. Compared with Jia \textit{et al.} \cite{jia2019content}, our model achieves comparative BD-rate reduction with much lower complexity. Finally, we also conduct the experiments on H.266/VVC and ablation studies to demonstrate the effectiveness of the model.
Our future work aims at further performance improvement of the learning-based filter in video coding.


\ifCLASSOPTIONcaptionsoff
  \newpage
\fi

\bibliographystyle{IEEEtran}
\bibliography{reference}
\end{document}